\documentclass[prd,aps,showpacs,twocolumn,a4paper,floatfix]{revtex4-1}

\usepackage{nicefrac}
\usepackage{mathrsfs}
\usepackage[english]{babel}
\usepackage{graphicx,psfrag}
\usepackage{amsmath,amsfonts,amssymb,amsthm}
\usepackage{subfigure}

\usepackage{hyperref}

\hypersetup{
     colorlinks   = true,
     citecolor    = blue
}

\hypersetup{colorlinks=true, citecolor=blue, linkcolor = magenta}

\usepackage{scalerel}

\usepackage{color}

\begin{document}
\preprint{APS/123-QED}

\title{Anisotropic Quark Stars with an Interacting Quark Equation of State}
\author{E. A. Becerra-Vergara}
\email{becerra.ea@gmail.com}
\author{Sindy Mojica}
\email{sindroc@gmail.com}
\author{F. D. Lora-Clavijo } 
\email{fadulora@uis.edu.co}
\author{Alejandro Cruz-Osorio} 
\email{alejandro.cruz@ext.uv.es}

\affiliation{
Grupo de Investigaci\'on en Relatividad y Gravitaci\'on\\ 
Escuela de F\'isica, Universidad Industrial de Santander\\
A. A. 678, Bucaramanga 680002, Colombia. \\
Departamento de Astronom\'ia y Astrof\'isica, Universitat de Val\'encia,\\
 Dr.  Moliner 50, 46100, Burjassot (Val\'encia), Spain.
}

\date{\today}
            
%%%%%%%%%%%%%%%%%%%%%
%%%   abstract   %%%%
%%%%%%%%%%%%%%%%%%%%%%
\begin{abstract}
A deep exploration of the parameter space that relates the interacting equation of state  with the bag constant B, and the interaction parameter a, is fundamental for the construction of diverse models of  quark stars. In particular, the anisotropy of quark stars with a well motivated quantum chromodynamics (QCD) equation of state is presented here. The contribution of the fourth order corrections parameter ($\mathrm{a}$) of the QCD perturbation on the radial and tangential pressure generate significant effects on the mass-radius relation and the stability of the quark star.  An adequate set of solutions for several values of the bag factor and the interaction parameter are used in order to calculate the relation between the mass, radius, density, compactness, and consequently the maximum masses and the stability. Therefore, while the more interactive quark solution lead to higher masses, the weak interaction among quarks give solutions similar to the  widely known MIT bag model.

%The anisotropy of quark stars with a well motivated quantum chromodynamics (QCD) equation of state is presented here. The contribution of the fourth order corrections parameter ($\mathrm{a}$) of the QCD perturbation on the radial and tangential pressure generate significant effects on the mass-radius relation and the stability of the quark star. Several models of anisotropic quark stars have been constructed, exploring the parameter space that defines the interacting equation of state, the bag constant $B$ and quark interaction parameter $a$, respectively. We present the mass-radius, interaction parameter-bag parameter and compactness diagrams including the recent outcomes on the mass and radius estimation of neutron stars. We found more massive stars for strong quark interaction and similar results to MIT bag model for weak quark interaction. 

%Also, a profile of solutions for the bag constant and the interaction parameter is obtained.

\end{abstract}

%%%%%%%%%%%%%%%%%%%%%
%%%   abstract   %%%%
%%%%%%%%%%%%%%%%%%%%%%

\maketitle

\section{Introduction}
%excat solutions 

Anisotropy of compact objects  \cite{Dev2000, Mak2003} is one of the main topics that have been studied in several astrophysical systems like boson stars \cite{Schunck2003}, gravastars \cite{Cattoen2005} and neutron stars \cite{Heintzmann1975}. Essentially, the anisotropy is presented as the difference between the radial and the tangential pressure $(P-P_{\perp})$ in the hydrostatic equilibrium equation which is obtained solving the Einstein field equations for the interior of the star. 

The ongoing knowledge about anisotropy could not have been possible without the statements settled by previous studies. For example, Ruderman \cite{Ruderman1972} theoretically showed that anisotropic effects could arise in stellar models, where nuclear matter reaches densities larger than $10^{15}g/cm^3$ due to the interactions that at this level are relativistic. Likewise, phase transitions \cite{Sokolov1980},\cite{Carter1998} between the inner core and the outer crust occur when the matter goes to a superfluid and superconductive state generating significant changes in the interior of a star. Also, the pion phase configuration \cite{Dev2000}, \cite{Sawyer1989}, and solid state configurations at densities of $10^{14} - 10^{15}$ orders of magnitude \cite{Canuto1974},\cite{Canuto1977}, and in other cases strong magnetic fields \cite{Yazadjiev2012} - \cite{Cardall2001} may produce anisotropies with observable consequences. These and other mechanisms producer of anisotropies can be found in \cite{Herrera1997}, and in the recent publication \cite{Dev2000}.

In order to measure the possible anisotropic effects, compact objects like neutron stars are taken as astrophysical laboratories to check how they are affected by this phenomena. In a pioneering work by Bowers $\&$ Liang \cite{Bowers1974}, the Einstein field equations for anisotropic spheres and incompressible matter were solved, and the effects of the anisotropy in the resulting maximum mass and red shift were discussed. Afterwards, Cosenza et al. \cite{Cosenza1981} presented a set of solutions with anisotropic sources based on known solutions for an isotropic matter. Later, numerical solutions \cite{Mak2003} exhibited a good agreement with the mass-radius relations, but also calculated the upper limit mass of a stable neutron star by taking an arbitrarily high value of the anisotropy \cite{Heintzmann1975}. 

Motivated by the current achievements in order to understand the processes that produce anisotropies, and the advances in nuclear physics that have shown the behavoiur of matter in the outer layers of a neutron star interior at certain densities, and in spite of the lack of knowledge about matter interactions at higher densities than the saturation nuclear matter density value, many theories have suggested that the neutron star core has \emph{exotic} \cite{Bombaci1997} constituents like hyperons, kaon condensates, or a deconfined phase of strange matter. Other theories suggest \cite{Glendenning1995} the existence of hybrid stars  made of hadronic matter mixed with quarks and a core purely made of quarks. However, quark stars (QS) can be generated by different processes, for instance, a core collapse after a supernova explosion \cite{Dai1995}, where the conversion of ordinary matter to quark matter in a deconfined core \cite{Rahaman2014}, \cite{Cheng1998} takes place. Other theories avocate for phase transitions that occur as consequence of the mass accretion in a low mass x- ray (LMXR) binary system \cite{Cheng1998b}. Therefore, there is no reason to ignore the existence of another type of compact objects apart from neutron stars. 

The usual EOS utilized to obtain solutions for quark stars is the well known MIT bag model \cite{chodos1974, Farhi1984} since it seems to be adequate to describe the behaviour of matter that is not governed by gravity but instead by the strong nuclear force. However, this EOS is not sufficiently powerful to characterize a system with interacting quarks or more complex structures. It is to expect that interactions among quarks must generate changes at the interior of the shell generating anisotropies that change the mass-radius relation and the gravitational redshift. Therefore, this also suggest that the equation of state can be constrained by considering the anisotropies at its interior produced by the mechanisms previously mentioned. So, in order to model anisotropic quark stars, several methods have been proposed. In \cite{Mak2002}, the anisotropy is modeled by taking two different expressions for the pressure, while the radial pressure is written as a lineal EOS, the tangential pressure is taken as a complex expression dependent of the radial coordinate, as a result this model yields to a mass-radius relation that exhibits values up to $3M_{\odot}$. On the other hand, a different approach given by a deterministic model \cite{Rahaman2014} used the MIT bag EOS, for this case observational evidence that contemplate the existence of strange stars is considered in order to make an interpolation function of the mass $m(r)$, where the obtained solutions are in agreement with Buchdahl model, which is explained with more detail in \cite{Buchdahl:59}. Another complementary work for the anisotropy of non-rotating strange stars and its effect in the usual physical observables is calculated in \cite{Maurya2016}, its aim is to test the stability of the model with a generalization of the Tolman-Oppenheimer-Volkoff equation and using the Herrera's cracking concept \cite{Herrera1992}. Meanwhile, other models are focused in the attempt to find a singularity-free solution of the Einstein equations \cite{Deb2017}, \cite{Deb2018} through the MIT bag model to obtain the mass-radius relation for different values of the bag constant. Additionally, in  \cite{Deb2017} the authors took the density profile given in \cite{Mak2002} and addressed their computations to obtain the total mass of the quark star. In this case, a general expression for the TOV equation was calculated, the stability of the system was evaluated with the Herrera's cracking concept, while the energy condition was satisfied. Another method \cite{Deb2018}, suggested a new model that uses the Homotopy Perturbation Method (HPM), in order to find a solution for spherically symmetric quark stars whose results were compared with quark stars candidates like \emph{CenX-3, VelaX-1}\cite{Rawls2011}, $4U 1820 - 30$ \cite{Guver2010}, $J1903+0327$ \cite{Freire2011}, $4U 1820-30$ \cite{Guver2010a}, $PSR J1614-2230$ \cite{Demorest2010}.

It is clear that following the current knowledge on the neutron stars and their layers, one should not discard the existence of more exotic objects, and although compact objects appear to be isotropic and homogeneous from the observations, it is impossible to think that their interiors are perfectly arranged to be considered as isotropic, since the nuclear phenomena that occur in the crust and the core are highly intense that certainty generate anisotropies and consequently produce changes on the mass-radius relation, as we shall see henceforth. In section (\ref{TOVEQ})  a QCD motivated EOS is introduced and the Tolman-Oppenheimer-Volkoff equations are obtained. In section (\ref{numerical}) the numerical details to obtain the mass-radius relation and the full set of solutions presented in section (\ref{masra}) are explained. Finally, the highlight results and further research are proposed in section (\ref{discussion}).

\section{The model}\label{TOVEQ}

\subsection{Quark Matter Equation of State} \label{QEOS}

In spite the fact that strange stars have not been directly observed yet, there are some candidates \cite{Li1995}, \cite{Bombaci1997}, \cite{Li1999} that could fit the EOS associated with this type of objects. Those candidates seem not to adjust their masses and radius to the neutron stars models, but by mean of a semi-empirical relation that calculates the strength of the magnetic field of a pulsar \cite{Li1999}, a range for the mass-radius relation is obtained giving a good description of strange star composers. A widely accepted quark star model is the MIT bag model that characterizes a degenerated Fermi gas of quarks \emph{up, down} and \emph{strange} \cite{Farhi1984}, 
\cite{chodos1974, Schmitt2010,Haensel07}. This is the simplest and more frequently used form to illustrate the interior a quark star \cite{chodos1974}. Nevertheless, quark stars are not such a simple objects that only depend on the bag constant $B$, indeed, this led to the construction of several models based on quantum chromodynamics (QCD) corrections of second and fourth order with the aim of giving an approximate characterization of confined quarks, like the presented in \cite{Flores2017}. This model not only includes the interactions among quarks, but also suggest the possibility of the existence of new matter states analogous to the superconductivity state BSC, that is a phase known as colour flavour locked phase (CFL). 

Following the EOS mentioned above \cite{Flores2017}, we consider homogeneously confined matter inside the star with 3-flavour neutral charge and a fixed strange quark mass $m_s$. But for simplicity, the superconductivity generated in the CFL phase is not taken into account, thus the EOS reduces to the expression \cite{Asbell2017}
\begin{equation} \label{Prad}
\begin{aligned}
P=&\dfrac{1}{3}\left(\epsilon-4B\right)-\dfrac{m_{s}^{2}}{3\pi}\sqrt{\dfrac{\epsilon-B}{\mathrm{a}}}\\
&+\dfrac{m_{s}^{4}}{12\pi^{2}}\left[1-\dfrac{1}{\mathrm{a}}+3\ln\left(\dfrac{8\pi}{3m_{s}^{2}}\sqrt{\dfrac{\epsilon-B}{\mathrm{a}}}\right)\right],
\end{aligned}
\end{equation}
where $\epsilon$ is the energy density of the homogeneously distributed quark matter, the quark strange mass is $m_{s}=100$ MeV \cite{Beringer2012}, $B$ is Bag constant whose values run between $B>57$ MeV/fm$^3$ and $B<92$ MeV/fm$^3$ which are determined by the stability condition with respect to iron nuclei for two-flavour and the three-flavour quark matter respectively. This implies that strange quark matter is absolutely stable for a range of energy densities of $57 < B < 92$ MeV/fm$^3$ \cite{Schmitt2010}. Finally, $\mathrm{a}$  is the parameter that comes from the QCD corrections on the pressure of the quark free Fermi sea, this parameter is related to the maximum mass of the star with values of $\approx 2M_{\odot}$ for $\mathrm{a}\approx 0.7$  \cite{Fraga2001}.

\subsection{Tolman-Oppenheimer-Volkoff Equations} \label{QEOS}

Let us consider an anisotropic fluid and a spherically symmetric spacetime, whose line element is given in terms of the components of the metric $g_{\alpha \beta}$  by
\begin{eqnarray}
\mathrm{d}s^2 &=& -c^{2}\alpha^{2} \mathrm{d}\mathrm{t}^{2} + \left(1 - \dfrac{2Gm}{c^{2}r}\right)^{-1} \mathrm{d}r^2 + r^2 \mathrm{d}\Omega, \label{ds2}
\end{eqnarray}
being $\alpha = \alpha(r)$, $m = m(r)$, $\mathrm{d}\Omega=\mathrm{d}\theta^2 + \sin^2\theta \ \mathrm{d}\phi^2 $, $G$ the gravitational constant and $c$ the speed of light. The energy momentum tensor can be written as
\begin{eqnarray}
T_{\alpha \beta} &=& (\epsilon + P_\perp) u_\alpha u_\beta + P_\perp g_{\alpha \beta} + (P - P_\perp) n_\alpha n_\beta, \label{T_desc}
\end{eqnarray}
where $P$ is the radial pressure and $P_\perp$ is the tangential pressure, which is explicitly expressed as follows 
%\begin{equation} \label{Ptan}
%\begin{aligned}
%P_\perp=&\dfrac{1}{3}\left(\epsilon-4B_{\perp}\right)-\dfrac{m_{s}^{2}}{3\pi}\sqrt{\dfrac{\epsilon-B_{\perp}}{\mathrm{a_{\perp}}}}\\
%&+\dfrac{m_{s}^{4}}{12\pi^{2}}\left[1-\dfrac{1}{\mathrm{a_{\perp}}}+3\ln\left(\dfrac{8\pi}{3m_{s}^{2}}\sqrt{\dfrac{\epsilon-B_{\perp}}{\mathrm{a_{\perp}}}}\right)\right],
%\end{aligned}
%\end{equation}
\begin{equation} \label{Ptan}
\begin{aligned}
P_\perp =  & P_c + \dfrac{1}{3}\left(\epsilon-4B_{\perp}\right)-\dfrac{m_{s}^{2}}{3\pi}\sqrt{\dfrac{\epsilon-B_{\perp}}{\mathrm{a_{\perp}}}}\\
&+\dfrac{m_{s}^{4}}{12\pi^{2}}\left[1-\dfrac{1}{\mathrm{a_{\perp}}}+3\ln\left(\dfrac{8\pi}{3m_{s}^{2}}\sqrt{\dfrac{\epsilon-B_{\perp}}{\mathrm{a_{\perp}}}}\right)\right]\\
&-\dfrac{1}{3}\left(\epsilon_c-4B_{\perp}\right)+\dfrac{m_{s}^{2}}{3\pi}\sqrt{\dfrac{\epsilon_c-B_{\perp}}{\mathrm{a_{\perp}}}}\\
&-\dfrac{m_{s}^{4}}{12\pi^{2}}\left[1-\dfrac{1}{\mathrm{a_{\perp}}}+3\ln\left(\dfrac{8\pi}{3m_{s}^{2}}\sqrt{\dfrac{\epsilon_c-B_{\perp}}{\mathrm{a_{\perp}}}}\right)\right],
%&+\dfrac{1}{3}\left(\epsilon_c-4B\right)-\dfrac{m_{s}^{2}}{3\pi}\sqrt{\dfrac{\epsilon_c-B}{\mathrm{a}}}\\
%&+\dfrac{m_{s}^{4}}{12\pi^{2}}\left[1-\dfrac{1}{\mathrm{a}}+3\ln\left(\dfrac{8\pi}{3m_{s}^{2}}\sqrt{\dfrac{\epsilon_c-B}{\mathrm{a}}}\right)\right],
\end{aligned}
\end{equation}
with $P_c$ and $\epsilon_c$ the radial pressure (\ref{Prad}) and the energy density, respectively, at the center of the star. From this expression it can be seen that the radial and tangential pressures are the same at $r=0$, that is the fluid is isotropic there. It is worth mentioning that $B_{\perp}$ and $\mathrm{a_{\perp}}$ parameters are the contributions on the tangential component of the pressure, and run in the same range of values as $B$ and $\mathrm{a}$. 
 
On the other hand, $u^\alpha u_\alpha = -1$ and $n^\alpha n_\alpha = 1$ such that 
\begin{eqnarray}
u^\alpha &=& \left[\dfrac{1}{c\alpha}, 0, 0, 0\right], \\
n^\alpha &=& \left[0 , \left(1 - \dfrac{2Gm}{c^{2}r}\right)^{1/2}, 0, 0\right].
\end{eqnarray}\
By solving the Einstein field equations and matter equations, a general expression for an anisotropic spherically symmetric compact star is obtained
\begin{eqnarray} 
\dfrac{\mathrm{d}m}{\mathrm{d}r} &=& 4\pi r^{2}\epsilon ,\\
\dfrac{\mathrm{d}P}{\mathrm{d}r} &=& -\dfrac{\left(\epsilon + \dfrac{P}{c^{2}}\right)\left( m + \dfrac{4\pi r^3 P}{c^2}\right)}{\dfrac{r^{2}}{G}\left(1-\dfrac{2Gm}{rc^2}\right)}- \frac{2}{r}(P-P_\perp), \label{hyd}\\
\dfrac{1}{\alpha}\dfrac{\mathrm{d}\alpha}{\mathrm{d}r} &=& \dfrac{G}{c^{2}r^{2}} \left( m + \dfrac{4\pi r^3 P}{c^2}\right)\left(1-\dfrac{2Gm}{rc^2}\right)^{-1}.
\end{eqnarray}
Notice that the Eq.(\ref{hyd}) is the only one that contains the contribution of the radial and tangential pressure by the difference $P - P_{\perp}$.

\section{Numerical Details} \label{numerical}

The numerical calculations presented in this paper were carried out by using the CAFE astrophysical code \cite{Lora2015}. All the simulations are computed using a fourth order Runge-Kutta integrator in a 1D spherical grid, which extends from $r = 0M$ to the outer domain boundary, $r_{max} = 100 M$.  In order to avoid the singular behavior at $r=0$, we follow the procedure showed in \cite{Guzman2012}, in which a Taylor expansion is made around this point. The resulting approximate regular equations are programmed for at least the first mesh point located at $r = \Delta r$, being $\Delta r$ the uniform spatial resolution of the grid. It is worth mentioning that the numerical simulations were carried out by using geometrised units, see appendix \ref{app:units}.

\section{Mass-Radius relation of an anisotropic quark star} \label{masra}

The mass-radius relation for spherically symmetric anisotropic quark star solutions were calculated for the case where the difference in the hydrostatic equation between the tangential and the radial pressure is non zero.  From Eq.(\ref{Prad}) and Eq.(\ref{Ptan}) the anisotropy occurs at the level of the tangential component of the pressure due to the spherical symmetry, so the radial composers $\mathrm{a}$ and $B$ are being fixed, while the tangential $\mathrm{a_{\perp}}$ and $B_{\perp}$ are varied.

\begin{figure}[h]
\centering
\includegraphics[scale=0.6]{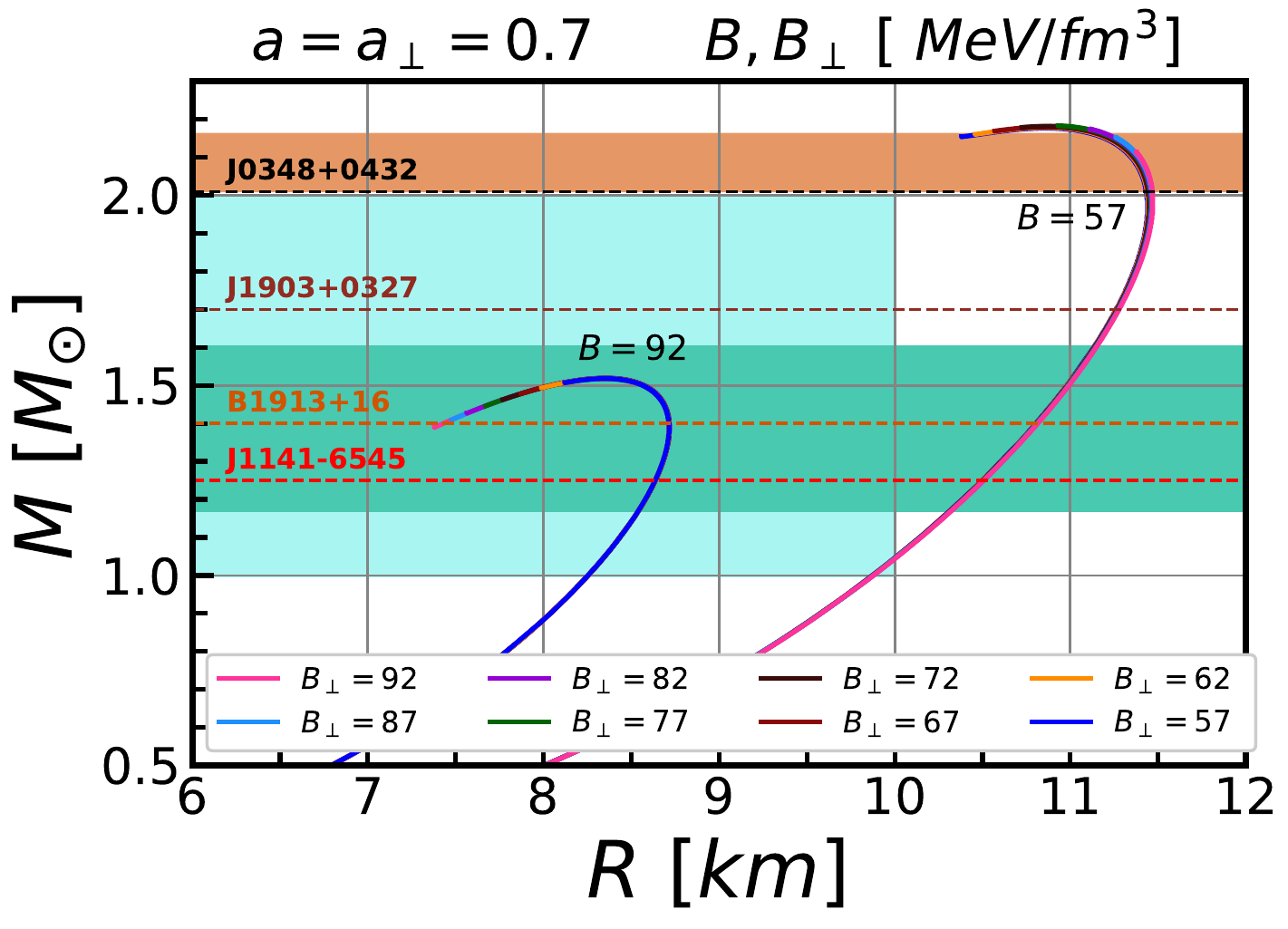}
\caption{Mass-radius relation for anisotropic quark stars with $a=a_{\perp}=0.7$. \label{fig1}}
\end{figure}

The first solution for the anisotropic quark star is displayed in Fig.(\ref{fig1}) presenting the mass-radius relation where the bag constant is set to $B=92$ $MeV/fm^{3}$. The parameters $\mathrm{a}=\mathrm{a_{\perp}}=0.7$ and $B_{\perp}$ take several values that cover a range among the isotropic solution \footnote{By taking the solutions of the Tolman-Oppenheimer-Volkoff equations given in section \ref{TOVEQ} reduce to the isotropic case when $(P-P_{\perp})=0$ in (Eq.(\ref{hyd}))}  i.e. $B_{\perp}=92$ $MeV/fm^{3}$, and the smallest value that the bag parameter can take, $B_{\perp}=57$ $MeV/fm^{3}$. As it is expected, $B_{\perp}$ do not produce any significant effect with respect to the isotropic case and all the solutions are overlapped in a unique curve. Furthermore, these solutions fall in the gravitational waves observation range  $1.17~M_{\odot} <M<1.6 ~M_{\odot}$  \cite{Abbott2017} (green region), and also in the upper limit mass  $2.01~M_{\odot} <M<2.16 ~M_{\odot}$ (orange region) recently found in \cite{Rezzolla2017}.

%\begin{widetext}
%\end{widetext}

%\begin{widetext}
\begin{figure}[h!]
\centering
\subfigure[]{\includegraphics[scale=0.6]{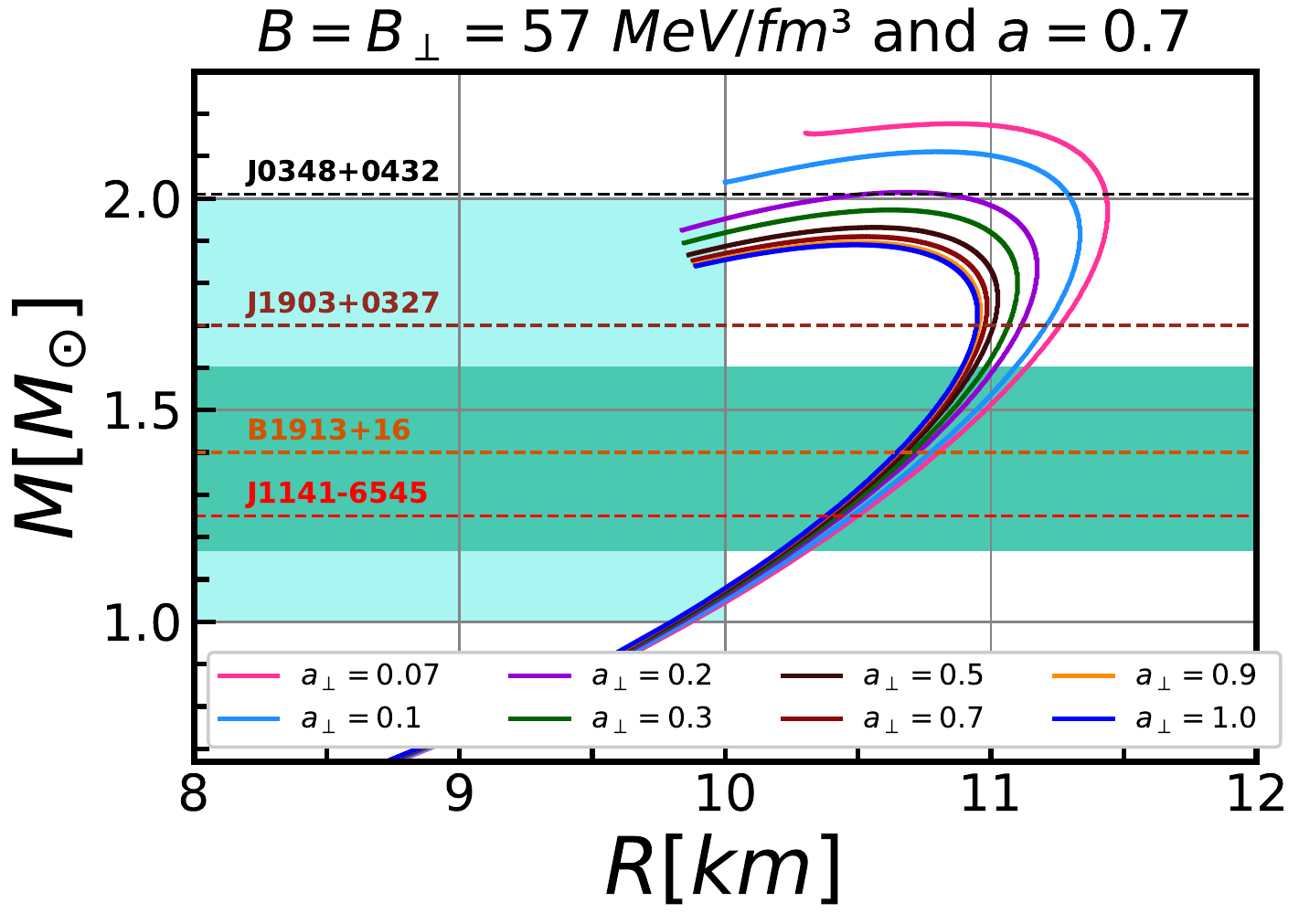}}\\
\subfigure[]{\includegraphics[scale=0.6]{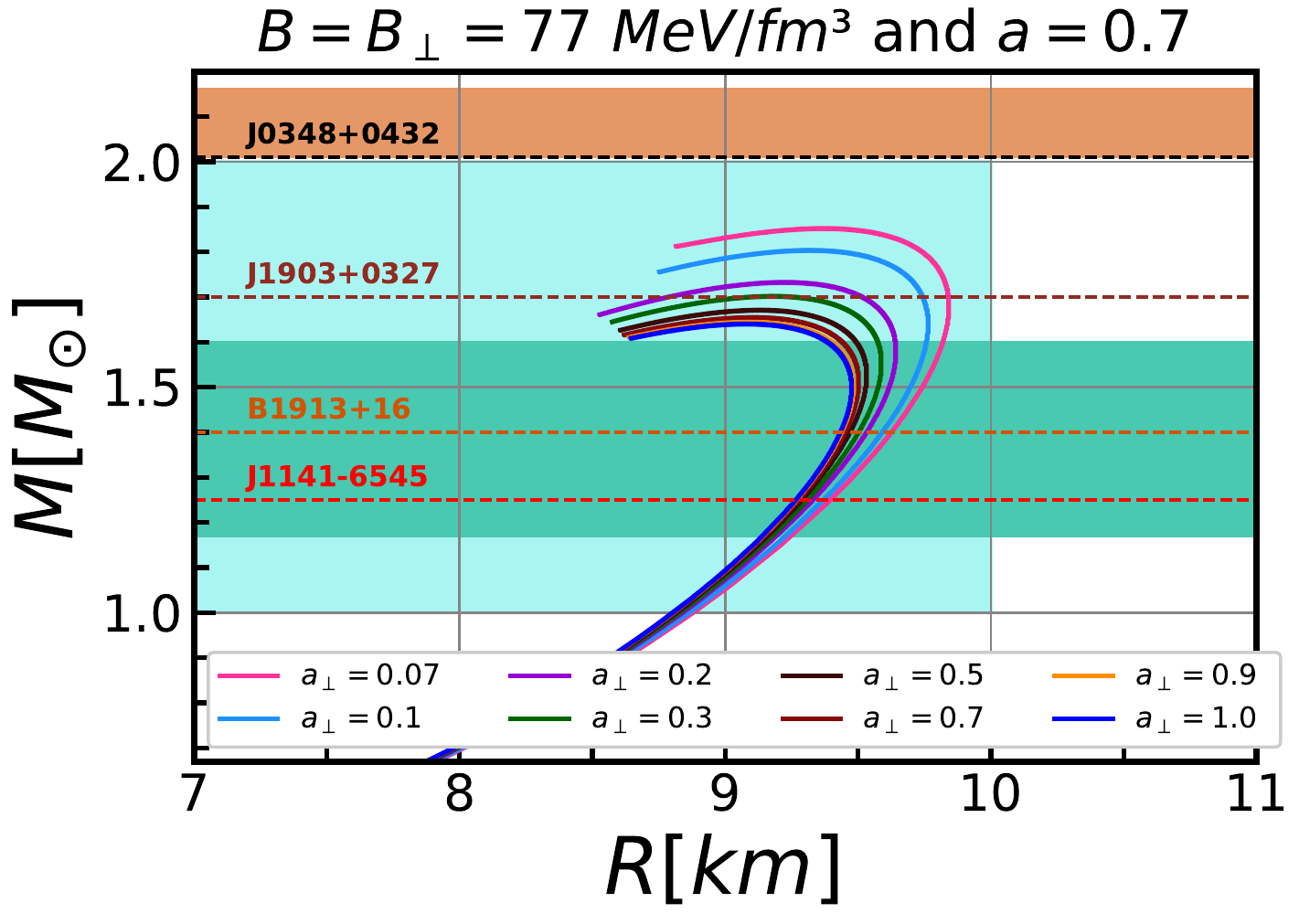}}\\
\subfigure[]{\includegraphics[scale=0.6]{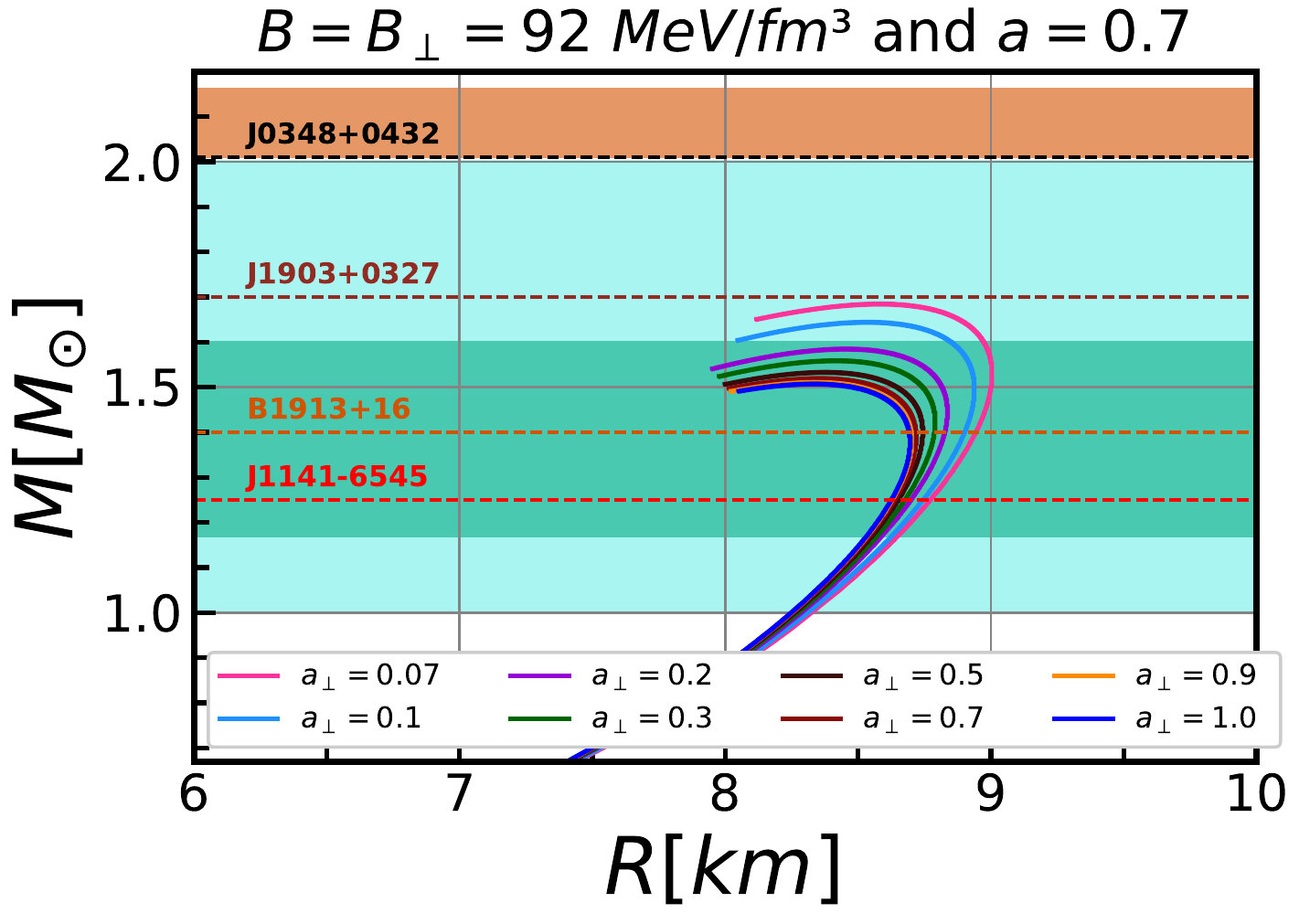}}
\caption{Mass-radius curve for anisotropic quark stars for (a) 
$B_{\perp}=57$, (b) $B_{\perp}=77$, and $B_{\perp}=92$ $MeV/fm^{3}$. The mass constriction (orange region) \cite{Rezzolla2017} and, the estimated mass from gravitational waves (green region) \cite{Abbott2017}, respectively.
\label{fig2}}
\end{figure}

\begin{figure}[h]
\centering
{\includegraphics[scale=0.6]{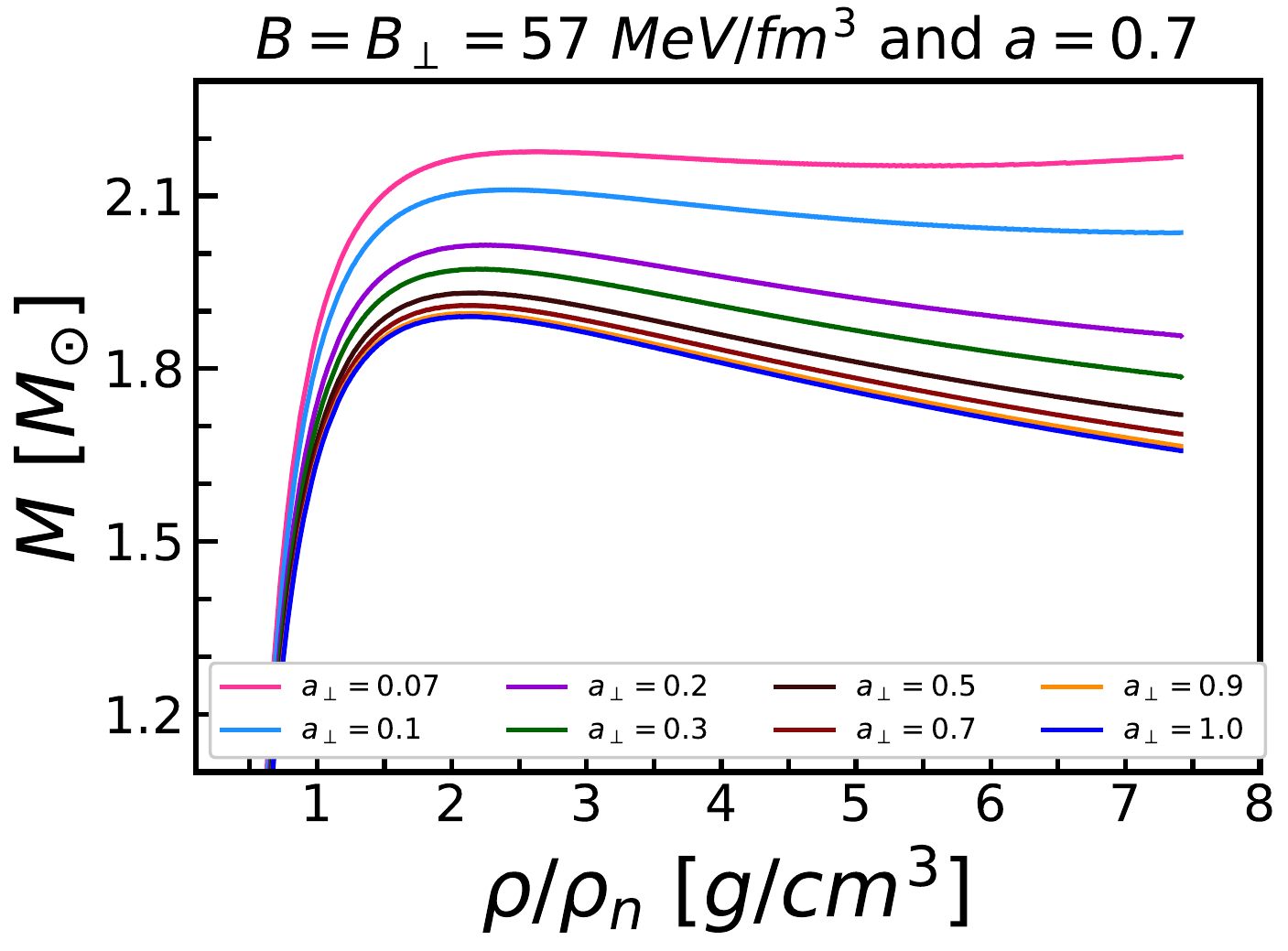}}
%\subfigure[]{\includegraphics[scale=0.6]{B77_rho.pdf}}
\caption{Mass-density relation for anisotropic quark star in where $\rho_n=2.4\times 10^{14}$ $g/cm^3$. \label{fig6}}
\end{figure}

%\begin{figure}[h]
%\centering
%\subfigure[]{\includegraphics[scale=0.6]{B92.pdf}}
%\subfigure[]{\includegraphics[scale=0.6]{B92_rho.pdf}}
%\caption{Mass-radius curve for anisotropic quark star.\label{fig7}}
%\end{figure}
%

%\begin{figure}[h]
%\centering
%\subfigure[]{\includegraphics[scale=0.6]{B77.pdf}}\subfigure[]{\includegraphics[scale=0.6]{compactness_B77.pdf} 
%}
%\caption{Compactness of anisotropic QS with $B_77$.\label{fig9}}
%\end{figure}
%
%\begin{figure}[h]
%\centering
%
%\subfigure[]{\includegraphics[scale=0.65]{B92.pdf}}\subfigure[]{\includegraphics[scale=0.65]{compactness_B92.pdf} 
%}
%\caption{Compactness of anisotropic QS with $B_92$.\label{fig10}}
%\end{figure}

%\end{widetext}

%\begin{widetext}
\begin{figure*}[h]
\begin{tabular}{c}
\subfigure[]{\includegraphics[scale=0.6]{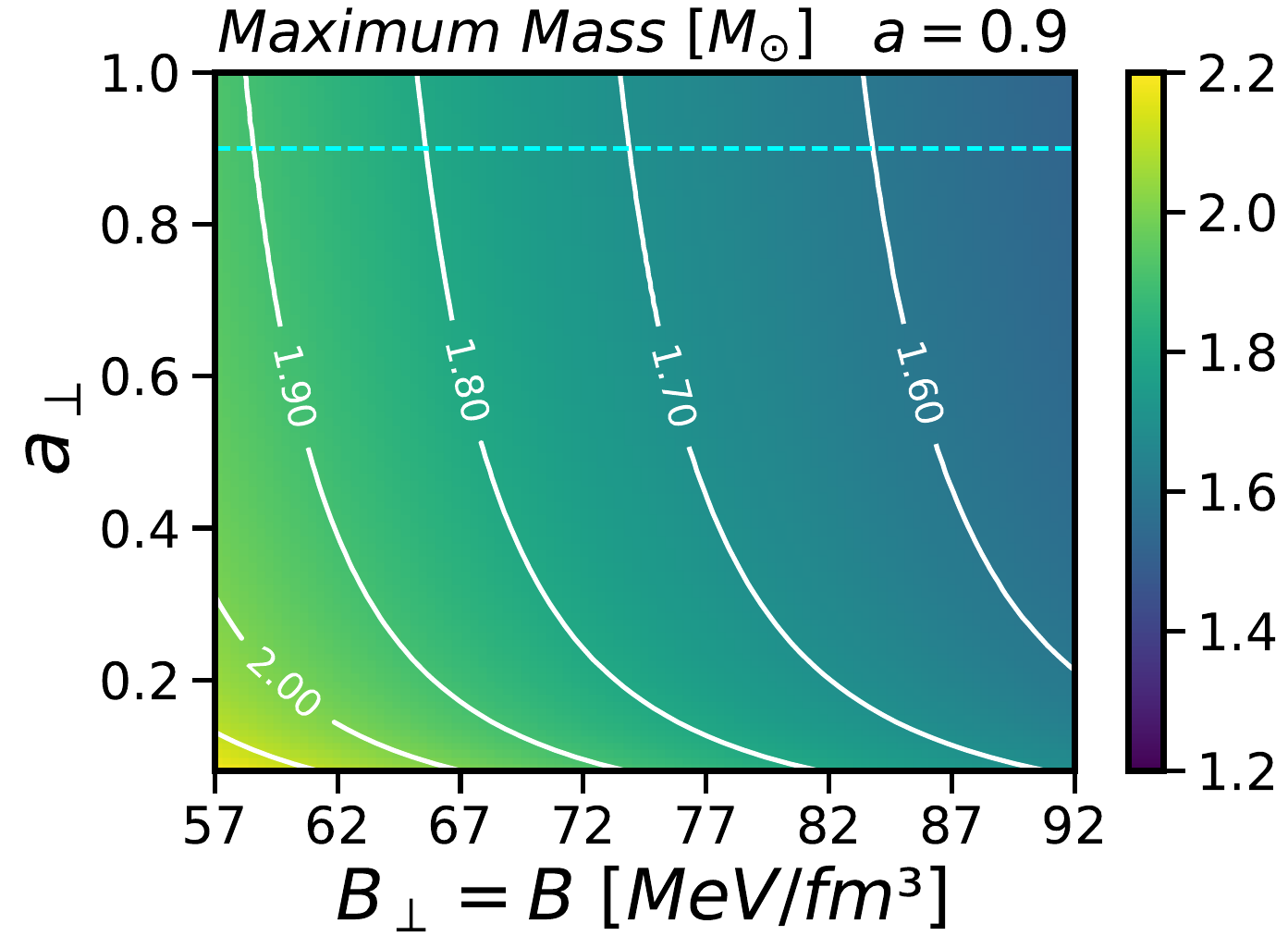}} \qquad  \subfigure[]{\includegraphics[scale=0.6]{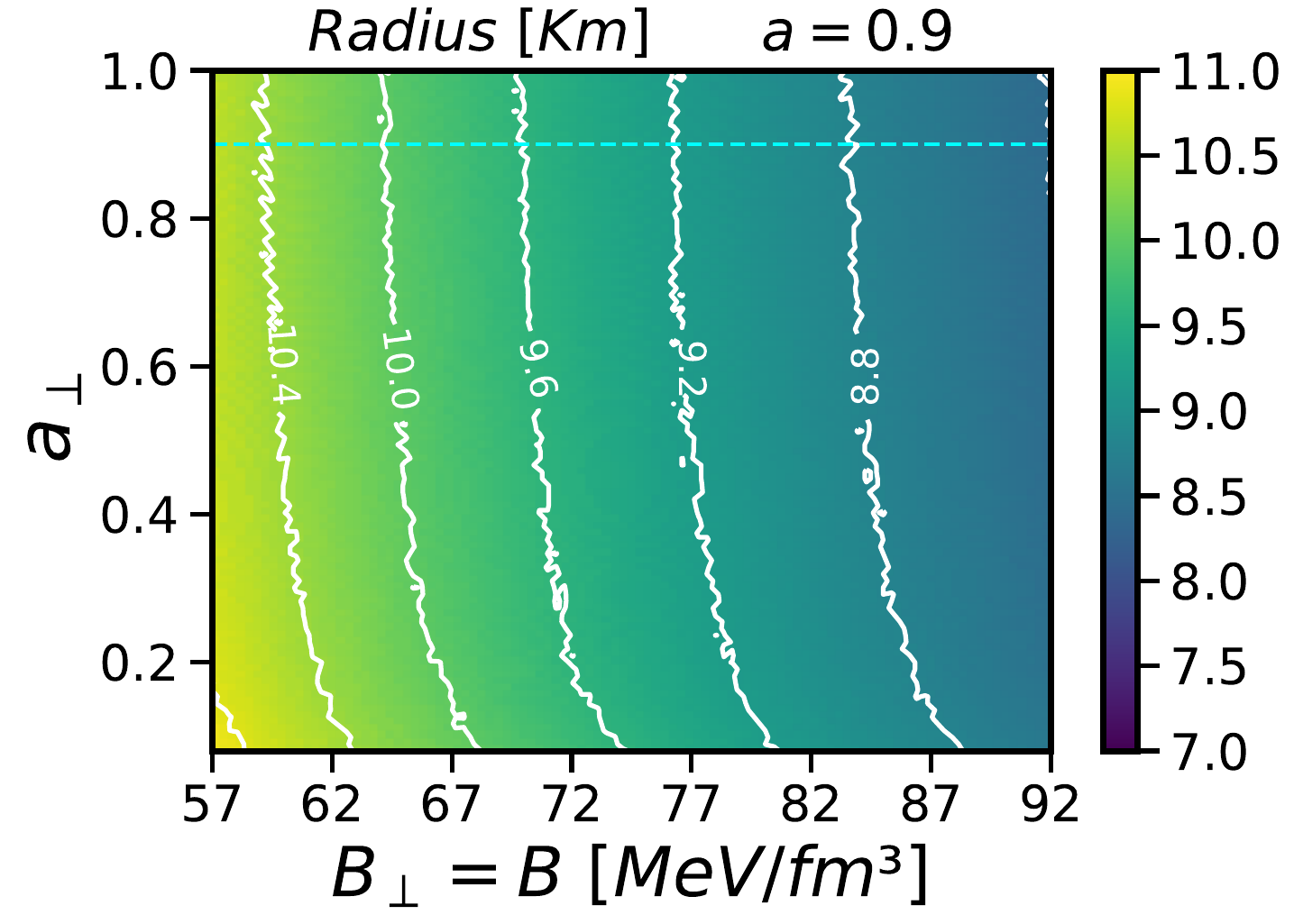} 
} \\
\subfigure[]{\includegraphics[scale=0.6]{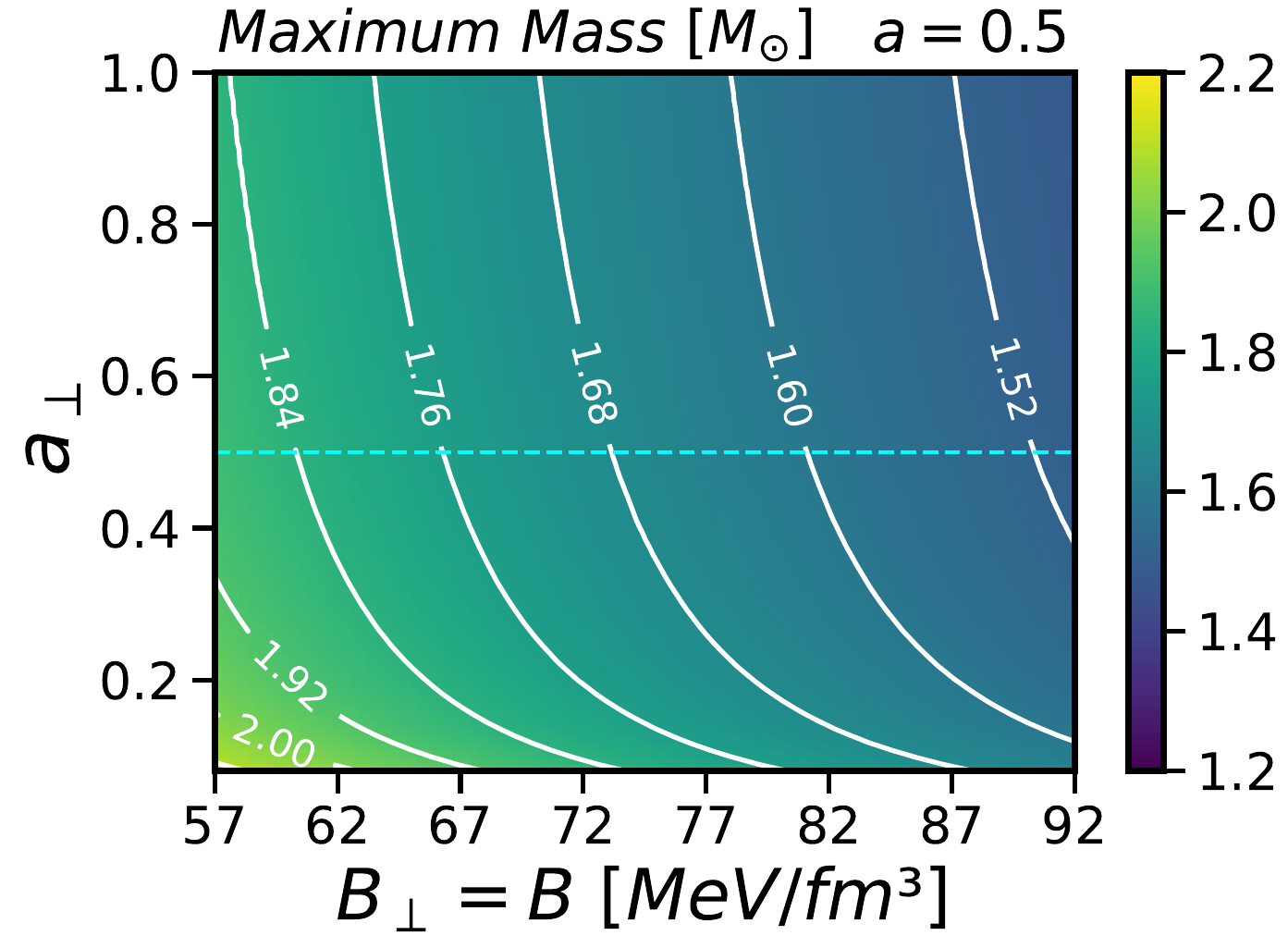}} \qquad \subfigure[]{\includegraphics[scale=0.6]{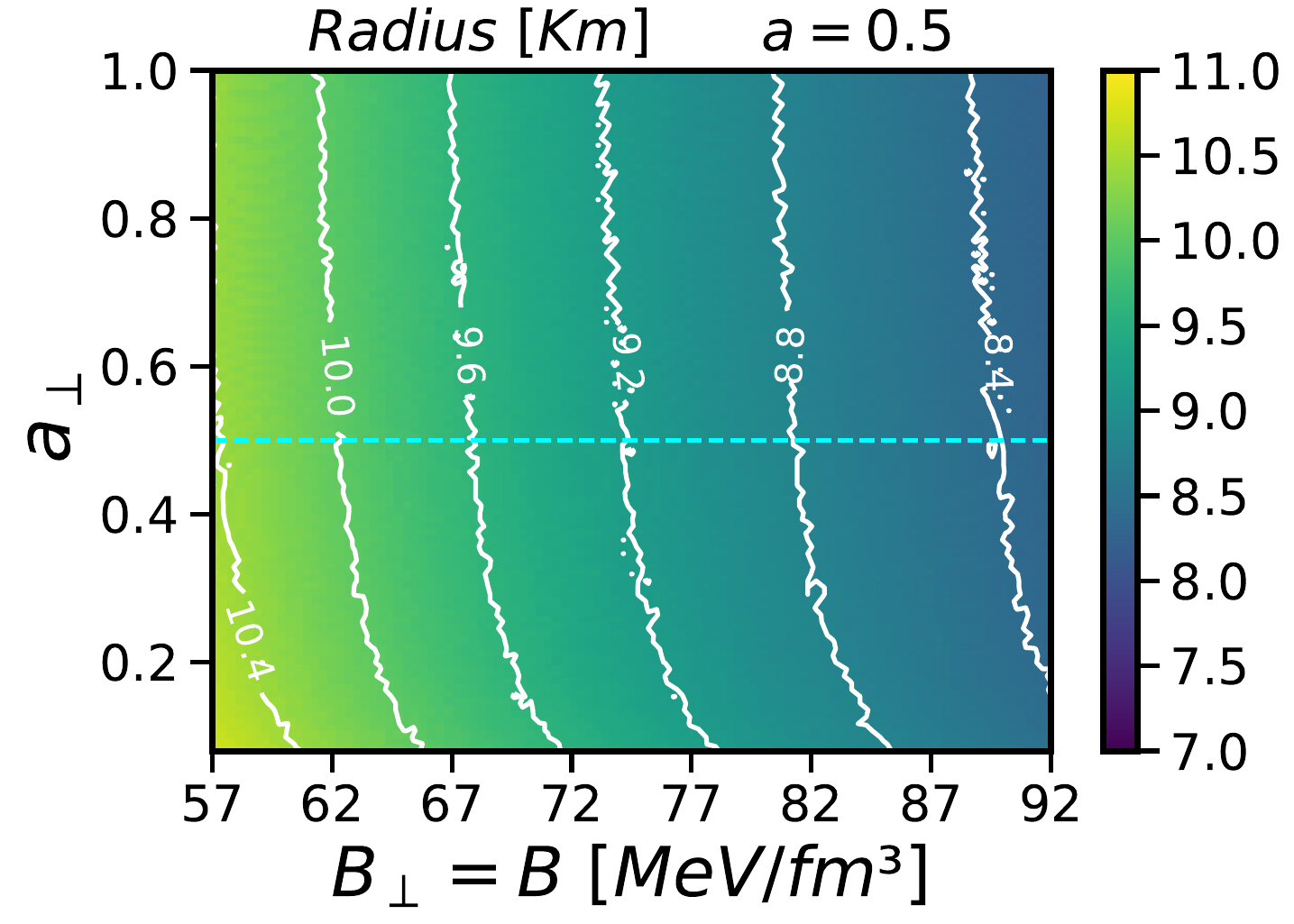} 
} \\
\subfigure[]{\includegraphics[scale=0.6]{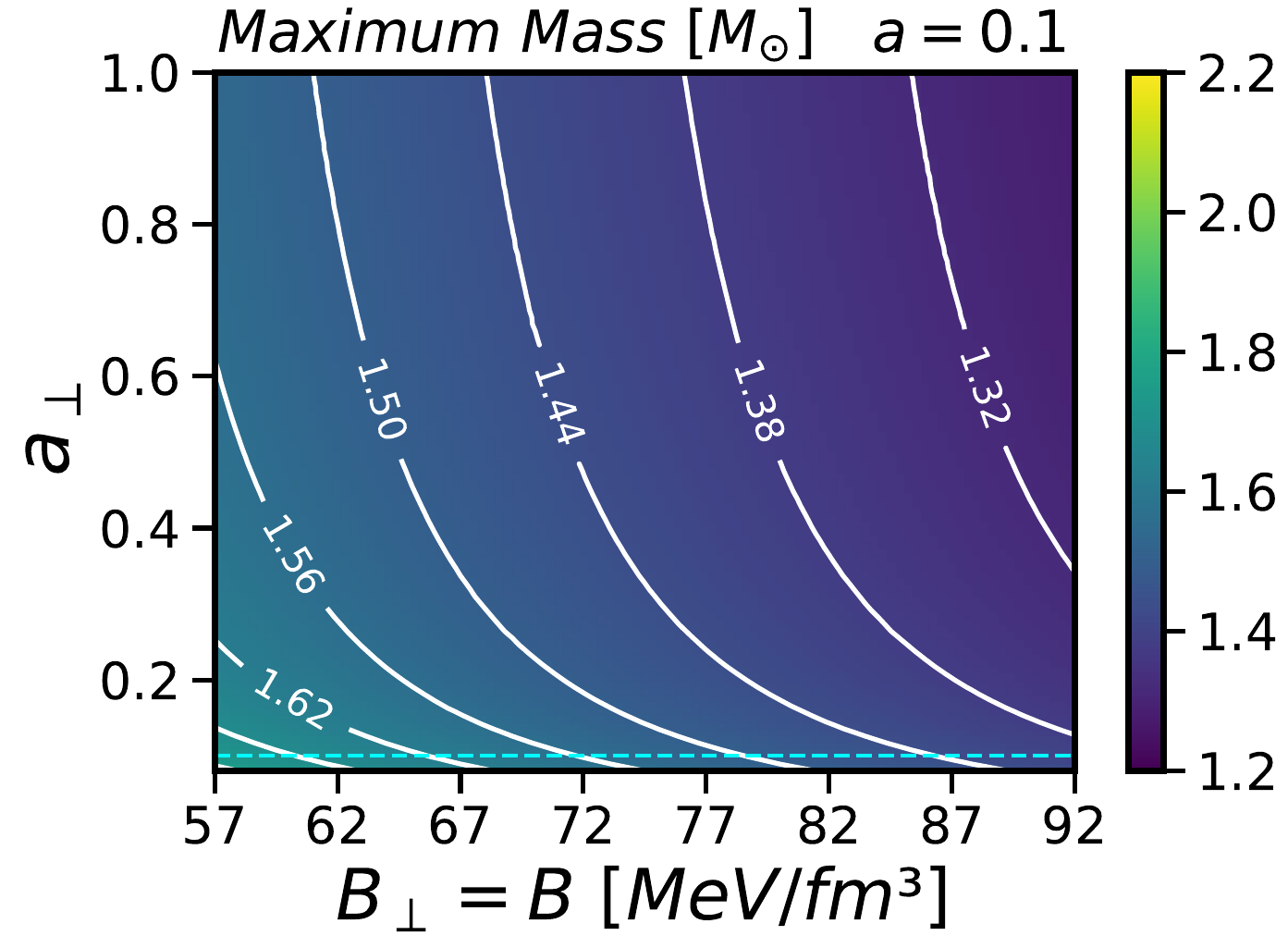}} \qquad \subfigure[]{\includegraphics[scale=0.6]{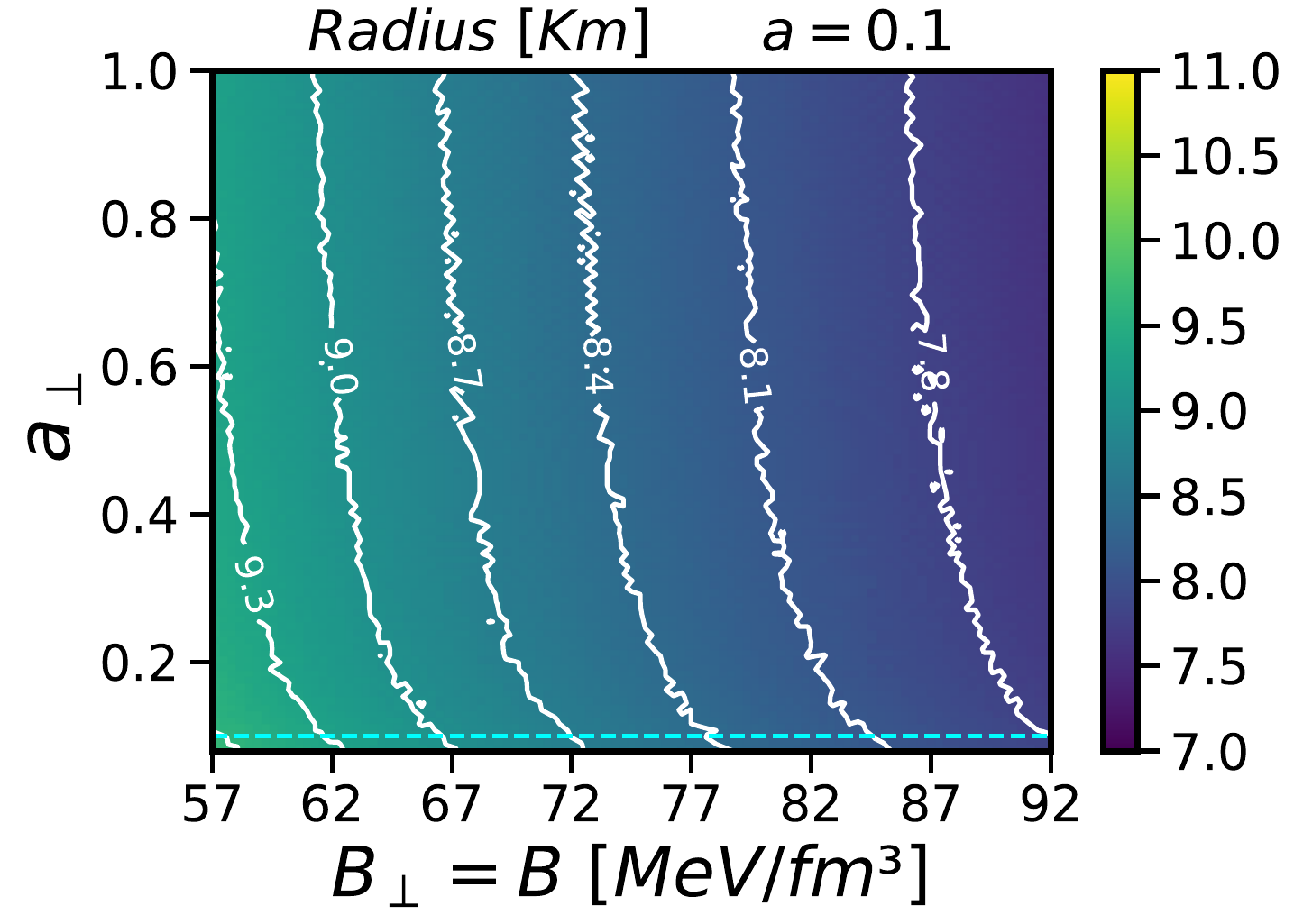}}

\end{tabular}

\caption{Maximum masses and their corresponding radii for the full range of values of  $\mathrm{a_{\perp}}$ and $\mathrm{B_{\perp}}$. The white lines are all QS in which the maximum mass and radius has the same value, and the cyan coulored lines are the isotropic cases ($\mathrm{a}=\mathrm{a_{\perp}}$). \label{fig10}}
\end{figure*}

In Fig. (\ref{fig2}a) the domain of solutions are around the static observational limit for neutron stars \cite{Antoniadis2013} (orange region), although the solutions do not exhibit a monotonic behavior, the maximum mass of the solution with anisotropic factor $\mathrm{a_{\perp}}=0.2$ fits with the line of the observed pulsar J0348+0432 \cite{Antoniadis2013}, which has a mass $\sim 2M_{\odot}$. By varying $\mathrm{a_{\perp}}$ it is evident that the maximum mass and radius increase their values up to approximately a $15\%$ and $4.5\%$, respectively. This is a crucial result because it is possible to restrict the EOS for a fixed value of the bag parameter. Notice that similar results are achieved for the intermediate (see Fig. \ref{fig2}b) and the upper limit (see Fig. \ref{fig2}c) of $B$. On the other hand, for 
$B=B_{\perp}=77$ the anisotropic parameter $\mathrm{a_{\perp}}=0.3$ (green solid line) fits with the observational mass for the pulsar J1903+0327,
%$B=B_{\perp}=77$ the anisotropic parameter $\mathrm{a_{\perp}}=0.3$ (green solid line) fits with the observational limit of the $2M_{\odot}$. 
moreover, the increment of the mass is approximate of a $12\%$ and the radius is around a $4.2\%$. Meanwhile, for $B=B_{\perp}=92$, most of the solutions are found in the region corresponding to pulsars  J1903+0327 and J11441-6545. Such masses are in agreement with estimations of gravitational waves from binary neutron star observations \cite{Abbott2017}.

Likewise, the mass-central density relation is plotted in Fig.(\ref{fig6}). It is evident that lowering the parameter $\mathrm{a_{\perp}}$, the solutions tend to be more stable, even though, they are not monotonically growing. In the same way, as the anisotropic factor is decreasing (i.e the interactions between quarks become stronger) the system is more stable. Notice that the stability is also influenced by the bag constant, in particular for $B=B_{\perp}=57$  $MeV/fm^{3}$ a saddle point seems to appear, although it does not happen for $B=B_{\perp}=77$ $MeV/fm^{3}$  and  $B=B_{\perp}=92$ $MeV/fm^{3}$. In a word, the anisotropy stabilizes the quark star.

On the other hand, a profile of solutions that covers the full range of values for $\mathrm{a_{\perp}}$ and $B_{\perp}$ is presented in Fig. (\ref{fig10}).  Although it is clear that $B$ do not generate anisotropies, it has a significant influence on the maximum masses and radius, since the increment of $B=B_{\perp}$ contributes to the diminution of these two observables, similar to the outcomes obtained for the isotropic case (cyan dashed line). It is evident that for less interacting quarks the maximum masses and their corresponding maximum radius have larger values, but note that for highly interacting quarks the maximum masses do not even reach the $2M_{\odot}$ constraint. Equally important, there is a sector where the solutions never reach solutions with $\mathrm{a_{\perp}}=1.0$, in fact, for $B_{\perp}$ below  $72 MeV/fm^{3}$ there are no maximum masses that reach the non-interacting quark limit. Certainly, this may be considered a strong restriction for the EOS.

Another key point are the results obtained to roughly describe the interior of quark stars candidates taking into account the maximum mass and maximum radius observations (white solid lines). Due to, it is possible to determine how much the quarks interact by restricting $B_{\perp}$, $\mathrm{a_{\perp}}$, and $\mathrm{a}$ for a given maximum mass and its corresponding radius. Nevertheless, it is not clear how does the anisotropy mechanism is produced by $\mathrm{a_{\perp}}$.

\section{Binding-Energy and compactness}

\begin{figure}[h!]
\centering
\includegraphics[scale=0.6]{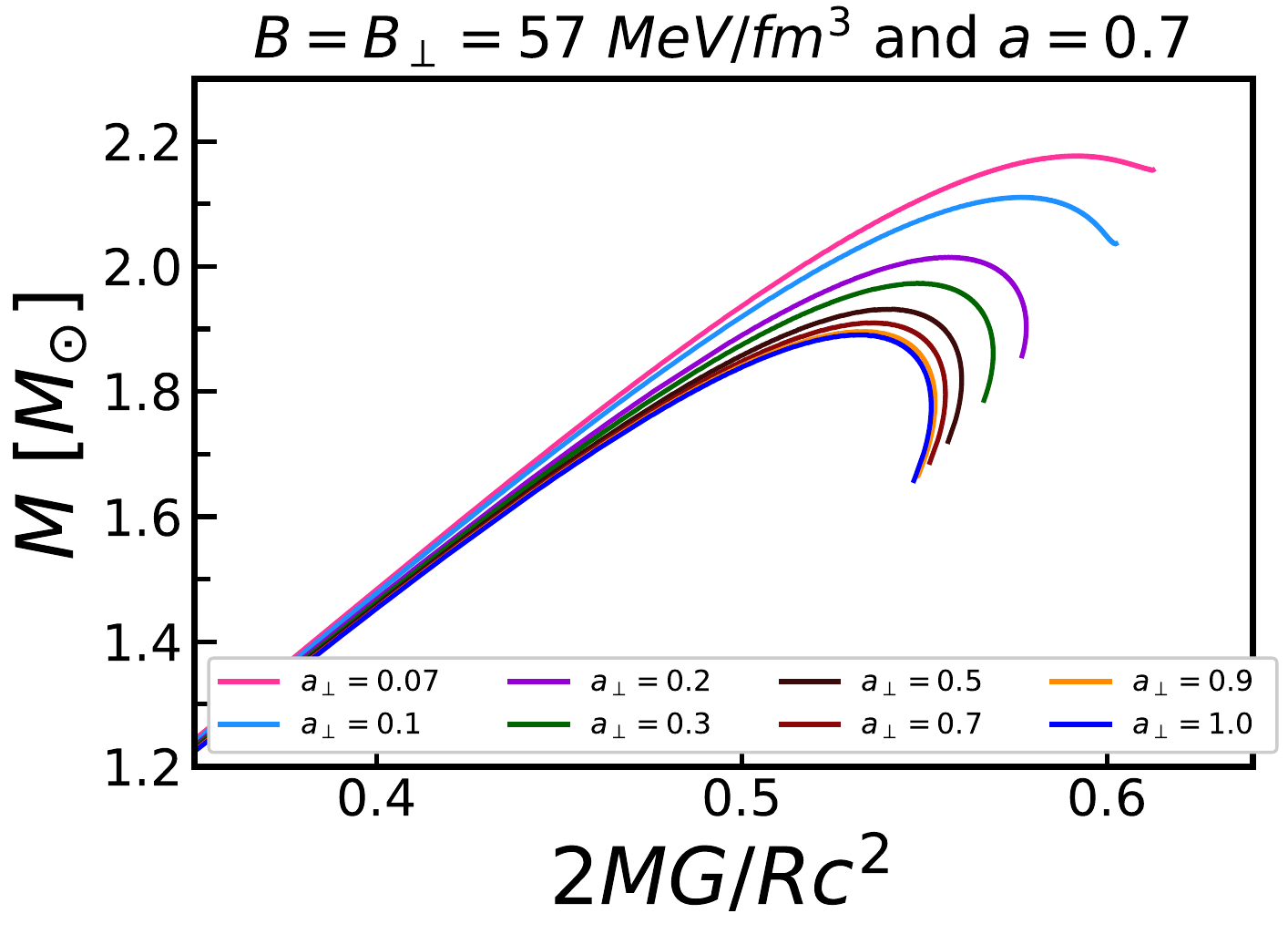} 
\caption{Compactness $\mathcal{C}$ of anisotropic QS with $B=57$.\label{fig8}}
\end{figure}

\begin{figure}[h!]
\centering
\subfigure[]{\includegraphics[scale=0.6]{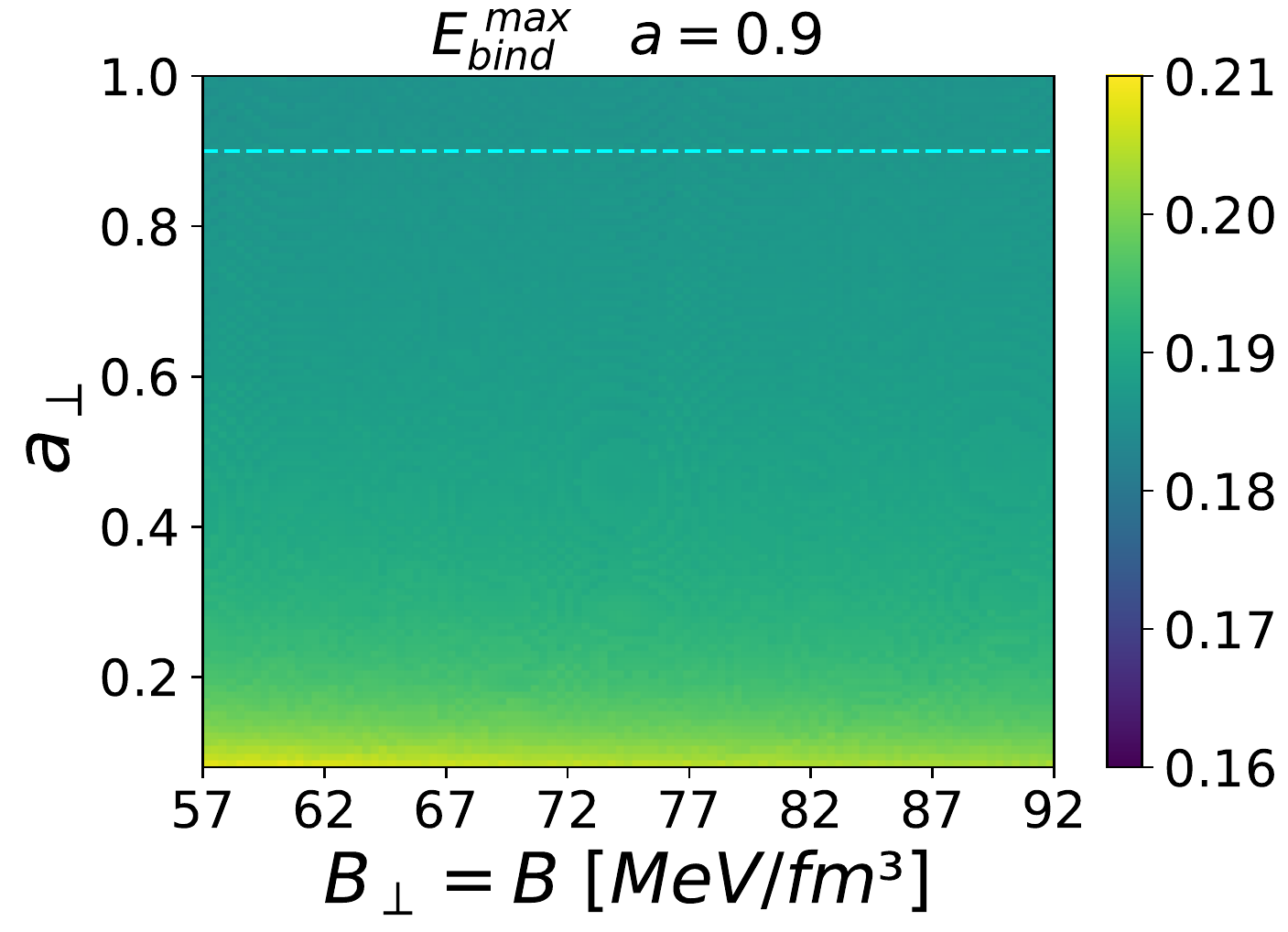}}\\
\subfigure[]{\includegraphics[scale=0.6]{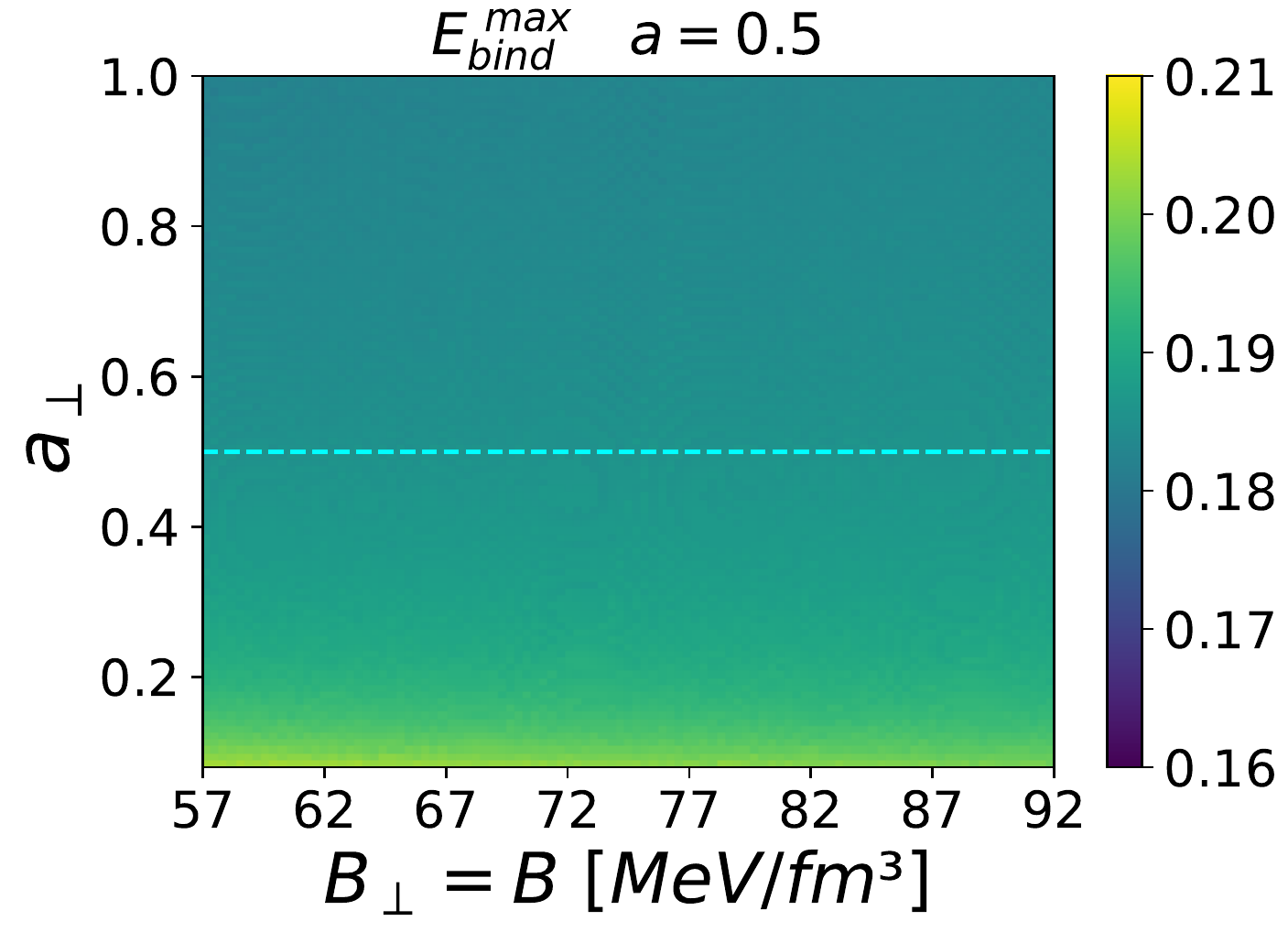}}\\
\subfigure[]{\includegraphics[scale=0.6]{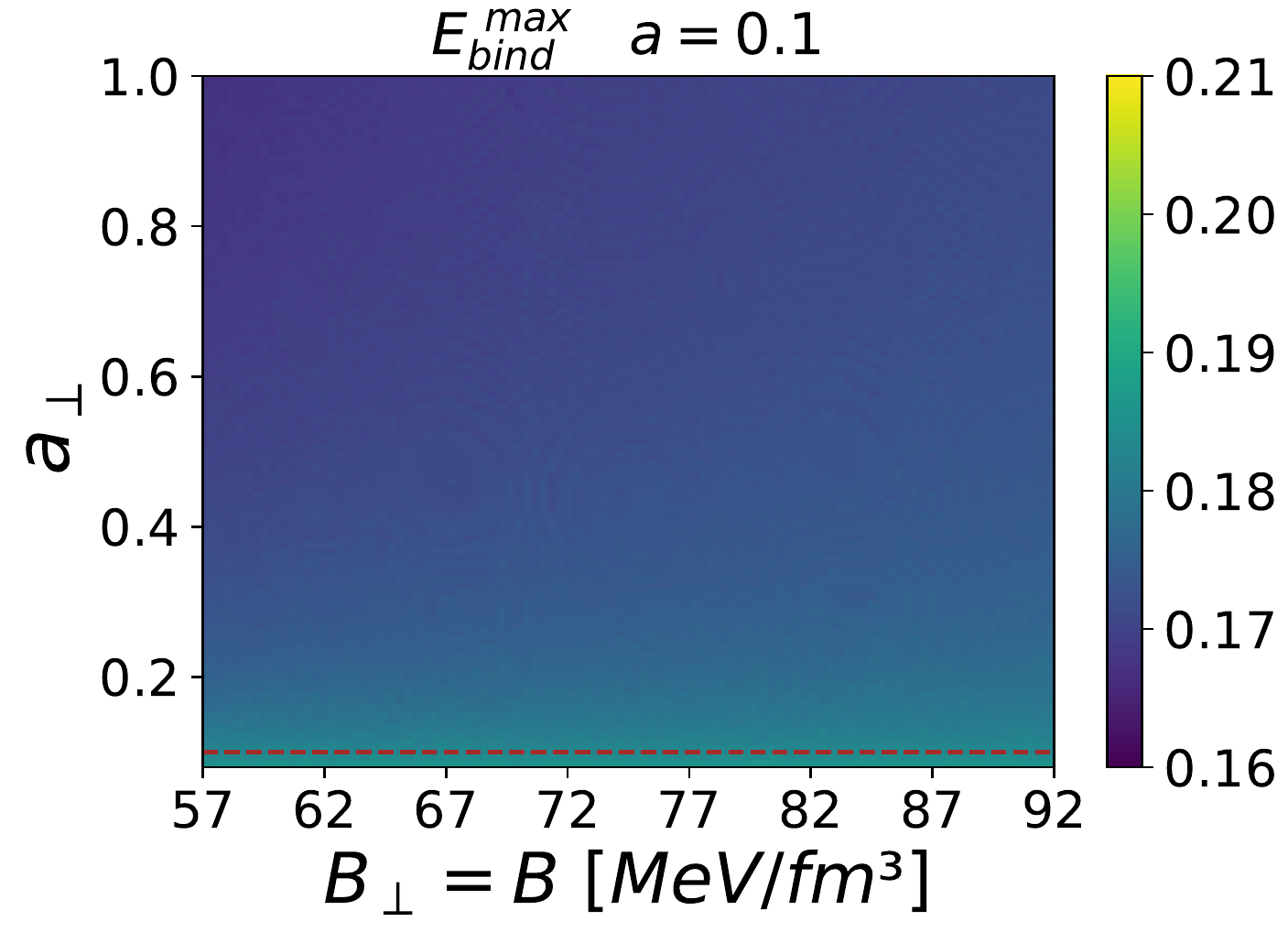}}
\caption{Binding energy profile for the full set of solutions of the perpendicular component of the interacting parameter and the bag parameter.\label{fig9}}
\end{figure}

The compactness $\mathcal{C}=2MG/Rc^2$ is sketched for a particular value of the bag parameter and a fixed $\mathrm{a}$ in Figure (\ref{fig8}), while the perpendicular contribution of the interacting term $\mathrm{a_{\perp}}$  run from  $1.0$ (blue solid line) to $0.07$ (pink solid line).  The comparison with respect to the isotropic case $\mathrm{a_{\perp}}=0.7$ (brown solid line), showst that the largest deviation of the compactness reaches up to a $14\%$. Additionally, we checked that the behaviour of the compactness is essentially the same for all the allowed values of $B=B_{\perp}$, i.e. it decreased as long as the quarks are less confined. 

On the other hand, the binding energy is explored for all the permitted values of the bag constant and the interacting parameter. It is evident from Fig. (\ref{fig9}) that the binding energy is larger for the lowest bag parameter $B$ and small values of anisotropic factor $\mathrm{a_{\perp}}$. We have observed that the binding energy decreases monotonically, but also it is not linear from $\mathrm{a}=0.9$ to $\mathrm{a}=0.1$. Although, the difference is not significantly large, the less interacting quarks have larger binding energy.

\section{discussions and conclusions}\label{discussion}

From the EOS (\ref{Prad}), it is clear that if the quark mass is zero, this reduces to the MIT bag model. On the other hand, the corrections introduced by the coefficient $\mathrm{a}$ produce similar results when there is no interaction among quark matter, but this is not clear at first sight due to the fact that $\mathrm{a}$ \cite{alford2005hybrid} has a parametrization that depends on the corrections of the pressure of the free-quark Fermi sea. In consequence, this leads to the condition that if $\mathrm{a} = 1$, then matter is made of free non-interacting quarks. For values between $0<\mathrm{a}<1$ the quark interactions play a significant role because they can increase the neutron stars mass up to $3-$times and this can explain the importance of considering a strange or exotic matter EOS in the neutron stars cores, supporting the idea that quark stars may have masses ($0.7 M_{\odot}$) smaller than the currently calculated for neutron stars\cite{schmitt2010dense}. Furthermore, the values of the interaction parameter that generate the observational contraint of $2 M_{\odot}$ are over $\mathrm{a} = 0.5$, as it can be seen in the green-yellow region of the figures (\ref{fig9} (a), (c)).   

%
%
%For $\mathrm{a} =1$ the model describes a non-interacting quark matter and it is equivalent to the MIT bag model \aco{We are sure?}, this gives masses of $0.7 M_{\odot}$ for neutron stars models \aco{in this case what is the value of $B$ and $B_{\perp}$? or maybe give us a range of masses?}. 

Considering only the simple MIT bag model to study the anisotropy may be a naive approximation, since many models agree with the idea of the existence of the interacting matter in the strange and quark stars interiors. Generally speaking, the value of $\mathrm{a_{\perp}}$ is a parameter associated to the QCD of quark stars, but also is a cut off of for their maximum and minimum masses. Therefore, these values can automatically restrict the quark star EOS through the anisotropy for a given $B$ constrained by the range of energies at which it runs ($57-92 MeV/fm^{3}$).

%tener valores de a de 1 me dan la masa m[]inima es el quivale a tener un sistema libre no interactuante de quarks (MIT), comprando con NS considerando el gas degenerado se obtienen masas por debajo (0.7) considerando interacciones nuclerares la masa alcanza la masa de NS. pero cuando se disminuye el a o sea mayor iteraccion entonces la masa de disminuye y me puede dar la restriccion de masa m]inuma de la estrella de quarks mayor interaccion posible entre quarks

 The results obtained above show that the effect of the bag parameter on the anisotropy is neglectable, this may be due to the geometry of the stellar structure itself, because it is restricted to be spherically symmetric and any change on $B$ will not affect this configuration. However, due to the interactions of the interior components of the quark stars are governed by the strong nuclear force, hence by the value of the interaction parameter, the major contribution to the anisotropy must come from $\mathrm{a_{\perp}}$. As larger $\mathrm{a_{\perp}}$ becomes the anisotropy produces less effect on the physical observables, since the quarks are almost free of interactions. However, for smaller values of $\mathrm{a_{\perp}}$, the interaction among quarks becomes larger, in consequence, the anisotropy generates significant changes of up to $15\%$ in the mass-radius relation.

%%%%%%%%%%%%%%%%%%%%%%%%%%%%%%

%%%%%%%%%%%%%%%%%%%%%%%%%%%%%%
\acknowledgments
%%%%%%%%%%%%%%%%%%%%%%%%%%%%%%
E. A. B-V. would like to thank the financial support from
COLCIENCIAS under the program Becas Doctorados Nacionales 
727 and Universidad Industrial de Santander. S. M. wants to thank the support from
the Postdoctoral Fellowship and the Grant No. 2416 provided by 
Universidad Industrial de Santander VIE-UIS. F.D.L-C was 
supported in part by VIE-UIS, under Grant No. 2314 and by COLCIENCIAS, 
Colombia, under Grant No. 8863. ACO gratefully acknowledges to CONACYT Postdoctoral Fellowship (291168, 291258). We thank
L. Rezzolla for the useful discussion. \\

\appendix
\section{}
\label{app:units}
%%%%%%%%%%%%%%%%%%%%%%%%%%%%%%

In order to use the adequate units for the calculations presented here, we utilize geometrised units ($G=c=1$) \cite{Rezzolla_book:2013} for the parameters of the equation of the state. Starting with the expression that relates de geometrised pressure and the c.g.s units
$$
P_{cgs} = 5.55173 \times 10^{38}\left(\frac{M_{\odot}}{M}\right)^2P_{geo},
 $$
and 
$$
1 \frac{MeV}{fm^3}=1.602176565 \times 10^{33} \frac{ergios}{cm^3}[=]\frac{din}{cm^2},
$$
the conversion factor for the bag constant given in by
$$
B_{geo}= \left(\frac{1.60218 \times 10^{33}}{5.55173  \times 10^{38}}\right)B_{[Mev/fm^3]}.
$$
Notice that the energy density $\epsilon$ has the same units as the bag constant.
Additionally the mass of the quark strange is 
$$
m_{s_{geo}}=\left(\frac{1}{5.55173 \times 10^{38}}\right)m_{s_{cgs}}.
$$

%%%%%%%%%%%%%%%%%%%%%%%%%%%%%%%%%%%%%%%%%%%%%%%%%%%%%%%%%%%%%
\bibliographystyle{apsrev}           
\bibliography{aeireferences}

\begin{thebibliography}{52}
\expandafter\ifx\csname natexlab\endcsname\relax\def\natexlab#1{#1}\fi
\expandafter\ifx\csname bibnamefont\endcsname\relax
  \def\bibnamefont#1{#1}\fi
\expandafter\ifx\csname bibfnamefont\endcsname\relax
  \def\bibfnamefont#1{#1}\fi
\expandafter\ifx\csname citenamefont\endcsname\relax
  \def\citenamefont#1{#1}\fi
\expandafter\ifx\csname url\endcsname\relax
  \def\url#1{\texttt{#1}}\fi
\expandafter\ifx\csname urlprefix\endcsname\relax\def\urlprefix{URL }\fi
\providecommand{\bibinfo}[2]{#2}
\providecommand{\eprint}[2][]{\url{#2}}

\bibitem[{\citenamefont{{Dev} and {Gleiser}}(2000)}]{Dev2000}
\bibinfo{author}{\bibfnamefont{K.}~\bibnamefont{{Dev}}} \bibnamefont{and}
  \bibinfo{author}{\bibfnamefont{M.}~\bibnamefont{{Gleiser}}},
  \bibinfo{journal}{ArXiv Astrophysics e-prints}  (\bibinfo{year}{2000}),
  \eprint{astro-ph/0012265}.

\bibitem[{\citenamefont{Mak and Harko}(2003)}]{Mak2003}
\bibinfo{author}{\bibfnamefont{M.~K.} \bibnamefont{Mak}} \bibnamefont{and}
  \bibinfo{author}{\bibfnamefont{T.}~\bibnamefont{Harko}},
  \bibinfo{journal}{Proceedings: Mathematical, Physical and Engineering
  Sciences} \textbf{\bibinfo{volume}{459}}, \bibinfo{pages}{pp. 393}
  (\bibinfo{year}{2003}), ISSN \bibinfo{issn}{13645021},
  \urlprefix\url{http://www.jstor.org/stable/3560113}.

\bibitem[{\citenamefont{Schunck and Mielke}(2003)}]{Schunck2003}
\bibinfo{author}{\bibfnamefont{F.~E.} \bibnamefont{Schunck}} \bibnamefont{and}
  \bibinfo{author}{\bibfnamefont{E.~W.} \bibnamefont{Mielke}},
  \bibinfo{journal}{Classical and Quantum Gravity}
  \textbf{\bibinfo{volume}{20}}, \bibinfo{pages}{R301} (\bibinfo{year}{2003}),
  \urlprefix\url{http://stacks.iop.org/0264-9381/20/i=20/a=201}.

\bibitem[{\citenamefont{{Cattoen} et~al.}(2005)\citenamefont{{Cattoen},
  {Faber}, and {Visser}}}]{Cattoen2005}
\bibinfo{author}{\bibfnamefont{C.}~\bibnamefont{{Cattoen}}},
  \bibinfo{author}{\bibfnamefont{T.}~\bibnamefont{{Faber}}}, \bibnamefont{and}
  \bibinfo{author}{\bibfnamefont{M.}~\bibnamefont{{Visser}}},
  \bibinfo{journal}{Class. Quantum Grav.} \textbf{\bibinfo{volume}{22}},
  \bibinfo{pages}{4189} (\bibinfo{year}{2005}), \eprint{arXiv:gr-qc/0505137}.

\bibitem[{\citenamefont{{Heintzmann} and {Hillebrandt}}(1975)}]{Heintzmann1975}
\bibinfo{author}{\bibfnamefont{H.}~\bibnamefont{{Heintzmann}}}
  \bibnamefont{and}
  \bibinfo{author}{\bibfnamefont{W.}~\bibnamefont{{Hillebrandt}}},
  \bibinfo{journal}{\aap} \textbf{\bibinfo{volume}{38}}, \bibinfo{pages}{51}
  (\bibinfo{year}{1975}).

\bibitem[{\citenamefont{{Ruderman}}(1972)}]{Ruderman1972}
\bibinfo{author}{\bibfnamefont{M.}~\bibnamefont{{Ruderman}}},
  \bibinfo{journal}{\araa} \textbf{\bibinfo{volume}{10}}, \bibinfo{pages}{427}
  (\bibinfo{year}{1972}).

\bibitem[{\citenamefont{{Sokolov}}(1980)}]{Sokolov1980}
\bibinfo{author}{\bibfnamefont{A.~I.} \bibnamefont{{Sokolov}}},
  \bibinfo{journal}{Soviet Journal of Experimental and Theoretical Physics}
  \textbf{\bibinfo{volume}{52}}, \bibinfo{pages}{575} (\bibinfo{year}{1980}).

\bibitem[{\citenamefont{{Carter} and {Langlois}}(1998)}]{Carter1998}
\bibinfo{author}{\bibfnamefont{B.}~\bibnamefont{{Carter}}} \bibnamefont{and}
  \bibinfo{author}{\bibfnamefont{D.}~\bibnamefont{{Langlois}}},
  \bibinfo{journal}{Nuclear Physics B} \textbf{\bibinfo{volume}{531}},
  \bibinfo{pages}{478} (\bibinfo{year}{1998}), \eprint{arXiv:gr-qc/9806024},
  \urlprefix\url{http://www.sciencedirect.com/science/article/pii/S0550321398004301}.

\bibitem[{\citenamefont{{Sawyer}}(1989)}]{Sawyer1989}
\bibinfo{author}{\bibfnamefont{R.~F.} \bibnamefont{{Sawyer}}},
  \bibinfo{journal}{Phys.~Rev.~D} \textbf{\bibinfo{volume}{39}},
  \bibinfo{pages}{3804} (\bibinfo{year}{1989}).

\bibitem[{\citenamefont{{Canuto}}(1974)}]{Canuto1974}
\bibinfo{author}{\bibfnamefont{V.}~\bibnamefont{{Canuto}}},
  \bibinfo{journal}{\araa} \textbf{\bibinfo{volume}{12}}, \bibinfo{pages}{167}
  (\bibinfo{year}{1974}).

\bibitem[{\citenamefont{{Canuto}}(1977)}]{Canuto1977}
\bibinfo{author}{\bibfnamefont{V.}~\bibnamefont{{Canuto}}}, in
  \emph{\bibinfo{booktitle}{Eighth Texas Symposium on Relativistic
  Astrophysics}}, edited by \bibinfo{editor}{\bibfnamefont{M.~D.}
  \bibnamefont{{Papagiannis}}} (\bibinfo{year}{1977}), vol.
  \bibinfo{volume}{302} of \emph{\bibinfo{series}{Annals of the New York
  Academy of Sciences}}, p. \bibinfo{pages}{514}.

\bibitem[{\citenamefont{{Yazadjiev}}(2012)}]{Yazadjiev2012}
\bibinfo{author}{\bibfnamefont{S.~S.} \bibnamefont{{Yazadjiev}}},
  \bibinfo{journal}{\prd} \textbf{\bibinfo{volume}{85}}, \bibinfo{eid}{044030}
  (\bibinfo{year}{2012}), \eprint{1111.3536}.

\bibitem[{\citenamefont{{Cardall} et~al.}(2001)\citenamefont{{Cardall},
  {Prakash}, and {Lattimer}}}]{Cardall2001}
\bibinfo{author}{\bibfnamefont{C.~Y.} \bibnamefont{{Cardall}}},
  \bibinfo{author}{\bibfnamefont{M.}~\bibnamefont{{Prakash}}},
  \bibnamefont{and} \bibinfo{author}{\bibfnamefont{J.~M.}
  \bibnamefont{{Lattimer}}}, \bibinfo{journal}{Astrophys. J.}
  \textbf{\bibinfo{volume}{554}}, \bibinfo{pages}{322} (\bibinfo{year}{2001}),
  \eprint{arXiv:astro-ph/0011148}.

\bibitem[{\citenamefont{{Herrera} and {Santos}}(1997)}]{Herrera1997}
\bibinfo{author}{\bibfnamefont{L.}~\bibnamefont{{Herrera}}} \bibnamefont{and}
  \bibinfo{author}{\bibfnamefont{N.~O.} \bibnamefont{{Santos}}},
  \bibinfo{journal}{\physrep} \textbf{\bibinfo{volume}{286}},
  \bibinfo{pages}{53} (\bibinfo{year}{1997}).

\bibitem[{\citenamefont{{Bowers} and {Liang}}(1974)}]{Bowers1974}
\bibinfo{author}{\bibfnamefont{R.~L.} \bibnamefont{{Bowers}}} \bibnamefont{and}
  \bibinfo{author}{\bibfnamefont{E.~P.~T.} \bibnamefont{{Liang}}},
  \bibinfo{journal}{\apj} \textbf{\bibinfo{volume}{188}}, \bibinfo{pages}{657}
  (\bibinfo{year}{1974}).

\bibitem[{\citenamefont{{Cosenza} et~al.}(1981)\citenamefont{{Cosenza},
  {Herrera}, {Esculpi}, and {Witten}}}]{Cosenza1981}
\bibinfo{author}{\bibfnamefont{M.}~\bibnamefont{{Cosenza}}},
  \bibinfo{author}{\bibfnamefont{L.}~\bibnamefont{{Herrera}}},
  \bibinfo{author}{\bibfnamefont{M.}~\bibnamefont{{Esculpi}}},
  \bibnamefont{and} \bibinfo{author}{\bibfnamefont{L.}~\bibnamefont{{Witten}}},
  \bibinfo{journal}{Journal of Mathematical Physics}
  \textbf{\bibinfo{volume}{22}}, \bibinfo{pages}{118} (\bibinfo{year}{1981}).

\bibitem[{\citenamefont{{Bombaci}}(1997)}]{Bombaci1997}
\bibinfo{author}{\bibfnamefont{I.}~\bibnamefont{{Bombaci}}},
  \bibinfo{journal}{\prc} \textbf{\bibinfo{volume}{55}}, \bibinfo{pages}{1587}
  (\bibinfo{year}{1997}).

\bibitem[{\citenamefont{{Glendenning}}(1995)}]{Glendenning1995}
\bibinfo{author}{\bibfnamefont{N.~K.} \bibnamefont{{Glendenning}}},
  \bibinfo{journal}{\apj} \textbf{\bibinfo{volume}{448}}, \bibinfo{pages}{797}
  (\bibinfo{year}{1995}).

\bibitem[{\citenamefont{{Dai} et~al.}(1995)\citenamefont{{Dai}, {Peng}, and
  {Lu}}}]{Dai1995}
\bibinfo{author}{\bibfnamefont{Z.}~\bibnamefont{{Dai}}},
  \bibinfo{author}{\bibfnamefont{Q.}~\bibnamefont{{Peng}}}, \bibnamefont{and}
  \bibinfo{author}{\bibfnamefont{T.}~\bibnamefont{{Lu}}},
  \bibinfo{journal}{\apj} \textbf{\bibinfo{volume}{440}}, \bibinfo{pages}{815}
  (\bibinfo{year}{1995}).

\bibitem[{\citenamefont{Rahaman et~al.}(2014)\citenamefont{Rahaman,
  Chakraborty, Kuhfittig, Shit, and Rahman}}]{Rahaman2014}
\bibinfo{author}{\bibfnamefont{F.}~\bibnamefont{Rahaman}},
  \bibinfo{author}{\bibfnamefont{K.}~\bibnamefont{Chakraborty}},
  \bibinfo{author}{\bibfnamefont{P.}~\bibnamefont{Kuhfittig}},
  \bibinfo{author}{\bibfnamefont{G.~C.} \bibnamefont{Shit}}, \bibnamefont{and}
  \bibinfo{author}{\bibfnamefont{M.}~\bibnamefont{Rahman}},
  \bibinfo{journal}{European Physical Journal C} \textbf{\bibinfo{volume}{74}},
  \bibinfo{eid}{3126} (\bibinfo{year}{2014}), \eprint{1406.4118}.

\bibitem[{\citenamefont{{Cheng} et~al.}(1998)\citenamefont{{Cheng}, {Dai}, and
  {Lu}}}]{Cheng1998}
\bibinfo{author}{\bibfnamefont{K.~S.} \bibnamefont{{Cheng}}},
  \bibinfo{author}{\bibfnamefont{Z.~G.} \bibnamefont{{Dai}}}, \bibnamefont{and}
  \bibinfo{author}{\bibfnamefont{T.}~\bibnamefont{{Lu}}},
  \bibinfo{journal}{International Journal of Modern Physics D}
  \textbf{\bibinfo{volume}{7}}, \bibinfo{pages}{139} (\bibinfo{year}{1998}).

\bibitem[{\citenamefont{{Cheng} and {Dai}}(1998)}]{Cheng1998b}
\bibinfo{author}{\bibfnamefont{K.~S.} \bibnamefont{{Cheng}}} \bibnamefont{and}
  \bibinfo{author}{\bibfnamefont{Z.~G.} \bibnamefont{{Dai}}},
  \bibinfo{journal}{Physical Review Letters} \textbf{\bibinfo{volume}{80}},
  \bibinfo{pages}{18} (\bibinfo{year}{1998}).

\bibitem[{\citenamefont{{Chodos} et~al.}(1974)\citenamefont{{Chodos}, {Jaffe},
  {Johnson}, {Thorn}, and {Weisskopf}}}]{chodos1974}
\bibinfo{author}{\bibfnamefont{A.}~\bibnamefont{{Chodos}}},
  \bibinfo{author}{\bibfnamefont{R.~L.} \bibnamefont{{Jaffe}}},
  \bibinfo{author}{\bibfnamefont{K.}~\bibnamefont{{Johnson}}},
  \bibinfo{author}{\bibfnamefont{C.~B.} \bibnamefont{{Thorn}}},
  \bibnamefont{and} \bibinfo{author}{\bibfnamefont{V.~F.}
  \bibnamefont{{Weisskopf}}}, \bibinfo{journal}{Phys. Rev. D}
  \textbf{\bibinfo{volume}{9}}, \bibinfo{pages}{3471} (\bibinfo{year}{1974}).

\bibitem[{\citenamefont{{Farhi} and {Jaffe}}(1984)}]{Farhi1984}
\bibinfo{author}{\bibfnamefont{E.}~\bibnamefont{{Farhi}}} \bibnamefont{and}
  \bibinfo{author}{\bibfnamefont{R.~L.} \bibnamefont{{Jaffe}}},
  \bibinfo{journal}{Phys. Rev. D} \textbf{\bibinfo{volume}{30}},
  \bibinfo{pages}{2379} (\bibinfo{year}{1984}).

\bibitem[{\citenamefont{{Mak} and {Harko}}(2002)}]{Mak2002}
\bibinfo{author}{\bibfnamefont{M.~K.} \bibnamefont{{Mak}}} \bibnamefont{and}
  \bibinfo{author}{\bibfnamefont{T.}~\bibnamefont{{Harko}}},
  \bibinfo{journal}{\cjaa} \textbf{\bibinfo{volume}{2}}, \bibinfo{pages}{248}
  (\bibinfo{year}{2002}).

\bibitem[{\citenamefont{Buchdahl}(1959)}]{Buchdahl:59}
\bibinfo{author}{\bibfnamefont{H.~A.} \bibnamefont{Buchdahl}},
  \bibinfo{journal}{Phys. Rev.} \textbf{\bibinfo{volume}{116}},
  \bibinfo{pages}{1027} (\bibinfo{year}{1959}).

\bibitem[{\citenamefont{{Maurya} et~al.}(2016)\citenamefont{{Maurya}, {Gupta},
  {Ray}, and {Deb}}}]{Maurya2016}
\bibinfo{author}{\bibfnamefont{S.~K.} \bibnamefont{{Maurya}}},
  \bibinfo{author}{\bibfnamefont{Y.~K.} \bibnamefont{{Gupta}}},
  \bibinfo{author}{\bibfnamefont{S.}~\bibnamefont{{Ray}}}, \bibnamefont{and}
  \bibinfo{author}{\bibfnamefont{D.}~\bibnamefont{{Deb}}},
  \bibinfo{journal}{European Physical Journal C} \textbf{\bibinfo{volume}{76}},
  \bibinfo{eid}{693} (\bibinfo{year}{2016}), \eprint{1607.05582}.

\bibitem[{\citenamefont{Herrera}(1992)}]{Herrera1992}
\bibinfo{author}{\bibfnamefont{L.}~\bibnamefont{Herrera}},
  \bibinfo{journal}{Physics Letters A} \textbf{\bibinfo{volume}{165}},
  \bibinfo{pages}{206} (\bibinfo{year}{1992}).

\bibitem[{\citenamefont{Deb et~al.}(2017)\citenamefont{Deb, Chowdhury, Ray,
  Rahaman, and Guha}}]{Deb2017}
\bibinfo{author}{\bibfnamefont{D.}~\bibnamefont{Deb}},
  \bibinfo{author}{\bibfnamefont{S.}~\bibnamefont{Chowdhury}},
  \bibinfo{author}{\bibfnamefont{S.}~\bibnamefont{Ray}},
  \bibinfo{author}{\bibfnamefont{F.}~\bibnamefont{Rahaman}}, \bibnamefont{and}
  \bibinfo{author}{\bibfnamefont{B.}~\bibnamefont{Guha}},
  \bibinfo{journal}{Annals of Physics} \textbf{\bibinfo{volume}{387}},
  \bibinfo{pages}{239} (\bibinfo{year}{2017}), \eprint{1606.00713}.

\bibitem[{\citenamefont{{Deb} et~al.}(2018)\citenamefont{{Deb}, {Roy
  Chowdhury}, {Ray}, and {Rahaman}}}]{Deb2018}
\bibinfo{author}{\bibfnamefont{D.}~\bibnamefont{{Deb}}},
  \bibinfo{author}{\bibfnamefont{S.}~\bibnamefont{{Roy Chowdhury}}},
  \bibinfo{author}{\bibfnamefont{S.}~\bibnamefont{{Ray}}}, \bibnamefont{and}
  \bibinfo{author}{\bibfnamefont{F.}~\bibnamefont{{Rahaman}}},
  \bibinfo{journal}{General Relativity and Gravitation}
  \textbf{\bibinfo{volume}{50}}, \bibinfo{eid}{112} (\bibinfo{year}{2018}),
  \eprint{1509.00401}.

\bibitem[{\citenamefont{Rawls et~al.}(2011)\citenamefont{Rawls, Orosz,
  McClintock, Torres, Bailyn, and Buxton}}]{Rawls2011}
\bibinfo{author}{\bibfnamefont{M.}~\bibnamefont{Rawls}},
  \bibinfo{author}{\bibfnamefont{J.}~\bibnamefont{Orosz}},
  \bibinfo{author}{\bibfnamefont{J.}~\bibnamefont{McClintock}},
  \bibinfo{author}{\bibfnamefont{M.}~\bibnamefont{Torres}},
  \bibinfo{author}{\bibfnamefont{C.}~\bibnamefont{Bailyn}}, \bibnamefont{and}
  \bibinfo{author}{\bibfnamefont{M.}~\bibnamefont{Buxton}},
  \bibinfo{journal}{\apj} \textbf{\bibinfo{volume}{730}}, \bibinfo{eid}{25}
  (\bibinfo{year}{2011}), \eprint{1101.2465}.

\bibitem[{\citenamefont{{G{\"u}ver} et~al.}(2010)\citenamefont{{G{\"u}ver},
  {{\"O}zel}, {Cabrera-Lavers}, and {Wroblewski}}}]{Guver2010}
\bibinfo{author}{\bibfnamefont{T.}~\bibnamefont{{G{\"u}ver}}},
  \bibinfo{author}{\bibfnamefont{F.}~\bibnamefont{{{\"O}zel}}},
  \bibinfo{author}{\bibfnamefont{A.}~\bibnamefont{{Cabrera-Lavers}}},
  \bibnamefont{and}
  \bibinfo{author}{\bibfnamefont{P.}~\bibnamefont{{Wroblewski}}},
  \bibinfo{journal}{\apj} \textbf{\bibinfo{volume}{712}}, \bibinfo{pages}{964}
  (\bibinfo{year}{2010}), \eprint{0811.3979}.

\bibitem[{\citenamefont{{Freire} et~al.}(2011)\citenamefont{{Freire}, {Bassa},
  {Wex}, {Stairs}, {Champion}, {Ransom}, {Lazarus}, {Kaspi}, {Hessels},
  {Kramer} et~al.}}]{Freire2011}
\bibinfo{author}{\bibfnamefont{P.~C.~C.} \bibnamefont{{Freire}}},
  \bibinfo{author}{\bibfnamefont{C.~G.} \bibnamefont{{Bassa}}},
  \bibinfo{author}{\bibfnamefont{N.}~\bibnamefont{{Wex}}},
  \bibinfo{author}{\bibfnamefont{I.~H.} \bibnamefont{{Stairs}}},
  \bibinfo{author}{\bibfnamefont{D.~J.} \bibnamefont{{Champion}}},
  \bibinfo{author}{\bibfnamefont{S.~M.} \bibnamefont{{Ransom}}},
  \bibinfo{author}{\bibfnamefont{P.}~\bibnamefont{{Lazarus}}},
  \bibinfo{author}{\bibfnamefont{V.~M.} \bibnamefont{{Kaspi}}},
  \bibinfo{author}{\bibfnamefont{J.~W.~T.} \bibnamefont{{Hessels}}},
  \bibinfo{author}{\bibfnamefont{M.}~\bibnamefont{{Kramer}}},
  \bibnamefont{et~al.}, \bibinfo{journal}{\mnras}
  \textbf{\bibinfo{volume}{412}}, \bibinfo{pages}{2763} (\bibinfo{year}{2011}),
  \eprint{1011.5809}.

\bibitem[{\citenamefont{G{\"u}ver et~al.}(2010)\citenamefont{G{\"u}ver,
  Wroblewski, Camarota, and {\"O}zel}}]{Guver2010a}
\bibinfo{author}{\bibfnamefont{T.}~\bibnamefont{G{\"u}ver}},
  \bibinfo{author}{\bibfnamefont{P.}~\bibnamefont{Wroblewski}},
  \bibinfo{author}{\bibfnamefont{L.}~\bibnamefont{Camarota}}, \bibnamefont{and}
  \bibinfo{author}{\bibfnamefont{F.}~\bibnamefont{{\"O}zel}},
  \bibinfo{journal}{\apj} \textbf{\bibinfo{volume}{719}}, \bibinfo{pages}{1807}
  (\bibinfo{year}{2010}), \eprint{1002.3825}.

\bibitem[{\citenamefont{{Demorest} et~al.}(2010)\citenamefont{{Demorest},
  {Pennucci}, {Ransom}, {Roberts}, and {Hessels}}}]{Demorest2010}
\bibinfo{author}{\bibfnamefont{P.~B.} \bibnamefont{{Demorest}}},
  \bibinfo{author}{\bibfnamefont{T.}~\bibnamefont{{Pennucci}}},
  \bibinfo{author}{\bibfnamefont{S.~M.} \bibnamefont{{Ransom}}},
  \bibinfo{author}{\bibfnamefont{M.~S.~E.} \bibnamefont{{Roberts}}},
  \bibnamefont{and} \bibinfo{author}{\bibfnamefont{J.~W.~T.}
  \bibnamefont{{Hessels}}}, \bibinfo{journal}{Nature}
  \textbf{\bibinfo{volume}{467}}, \bibinfo{pages}{1081} (\bibinfo{year}{2010}),
  \eprint{1010.5788}.

\bibitem[{\citenamefont{{Li} et~al.}(1995)\citenamefont{{Li}, {Dai}, and
  {Wang}}}]{Li1995}
\bibinfo{author}{\bibfnamefont{X.-D.} \bibnamefont{{Li}}},
  \bibinfo{author}{\bibfnamefont{Z.-G.} \bibnamefont{{Dai}}}, \bibnamefont{and}
  \bibinfo{author}{\bibfnamefont{Z.-R.} \bibnamefont{{Wang}}},
  \bibinfo{journal}{\aap} \textbf{\bibinfo{volume}{303}}, \bibinfo{pages}{L1}
  (\bibinfo{year}{1995}).

\bibitem[{\citenamefont{Li et~al.}(1999)\citenamefont{Li, Bombaci, Dey, Dey,
  and van~den Heuvel}}]{Li1999}
\bibinfo{author}{\bibfnamefont{X.-D.} \bibnamefont{Li}},
  \bibinfo{author}{\bibfnamefont{I.}~\bibnamefont{Bombaci}},
  \bibinfo{author}{\bibfnamefont{M.}~\bibnamefont{Dey}},
  \bibinfo{author}{\bibfnamefont{J.}~\bibnamefont{Dey}}, \bibnamefont{and}
  \bibinfo{author}{\bibfnamefont{E.~P.~J.} \bibnamefont{van~den Heuvel}},
  \bibinfo{journal}{Physical Review Letters} \textbf{\bibinfo{volume}{83}},
  \bibinfo{pages}{3776} (\bibinfo{year}{1999}), \eprint{hep-ph/9905356}.

\bibitem[{\citenamefont{{Schmitt}}(2010)}]{Schmitt2010}
\bibinfo{editor}{\bibfnamefont{A.}~\bibnamefont{{Schmitt}}}, ed.,
  \emph{\bibinfo{title}{{Dense Matter in Compact Stars}}}, vol.
  \bibinfo{volume}{811} of \emph{\bibinfo{series}{Lecture Notes in Physics,
  Berlin Springer Verlag}} (\bibinfo{year}{2010}), \eprint{1001.3294}.

\bibitem[{\citenamefont{Haensel et~al.}(2007)\citenamefont{Haensel, Potekhin,
  and Yakovlev}}]{Haensel07}
\bibinfo{author}{\bibfnamefont{P.}~\bibnamefont{Haensel}},
  \bibinfo{author}{\bibfnamefont{A.}~\bibnamefont{Potekhin}}, \bibnamefont{and}
  \bibinfo{author}{\bibfnamefont{D.}~\bibnamefont{Yakovlev}},
  \emph{\bibinfo{title}{Neutron Stars 1: Equation of State and Structure}},
  Astrophysics and Space Science Library (\bibinfo{publisher}{Springer New
  York}, \bibinfo{year}{2007}), ISBN \bibinfo{isbn}{9780387473017},
  \urlprefix\url{https://books.google.com.co/books?id=fgj\_TZ06niYC}.

\bibitem[{\citenamefont{Flores et~al.}(2017)\citenamefont{Flores, Hall, and
  Jaikumar}}]{Flores2017}
\bibinfo{author}{\bibfnamefont{C.~V.} \bibnamefont{Flores}},
  \bibinfo{author}{\bibfnamefont{Z.~B.} \bibnamefont{Hall}}, \bibnamefont{and}
  \bibinfo{author}{\bibfnamefont{P.}~\bibnamefont{Jaikumar}},
  \bibinfo{journal}{Phys. Rev. C} \textbf{\bibinfo{volume}{96}},
  \bibinfo{pages}{065803} (\bibinfo{year}{2017}),
  \urlprefix\url{https://link.aps.org/doi/10.1103/PhysRevC.96.065803}.

\bibitem[{\citenamefont{{Asbell} and {Jaikumar}}(2017)}]{Asbell2017}
\bibinfo{author}{\bibfnamefont{J.}~\bibnamefont{{Asbell}}} \bibnamefont{and}
  \bibinfo{author}{\bibfnamefont{P.}~\bibnamefont{{Jaikumar}}}, in
  \emph{\bibinfo{booktitle}{Journal of Physics Conference Series}}
  (\bibinfo{year}{2017}), vol. \bibinfo{volume}{861} of
  \emph{\bibinfo{series}{Journal of Physics Conference Series}}, p.
  \bibinfo{pages}{012029}, \eprint{1702.05691}.

\bibitem[{\citenamefont{{Beringer} et~al.}(2012)\citenamefont{{Beringer},
  {Arguin}, {Barnett}, {Copic}, {Dahl}, {Groom}, {Lin}, {Lys}, {Murayama},
  {Wohl} et~al.}}]{Beringer2012}
\bibinfo{author}{\bibfnamefont{J.}~\bibnamefont{{Beringer}}},
  \bibinfo{author}{\bibfnamefont{J.-F.} \bibnamefont{{Arguin}}},
  \bibinfo{author}{\bibfnamefont{R.~M.} \bibnamefont{{Barnett}}},
  \bibinfo{author}{\bibfnamefont{K.}~\bibnamefont{{Copic}}},
  \bibinfo{author}{\bibfnamefont{O.}~\bibnamefont{{Dahl}}},
  \bibinfo{author}{\bibfnamefont{D.~E.} \bibnamefont{{Groom}}},
  \bibinfo{author}{\bibfnamefont{C.-J.} \bibnamefont{{Lin}}},
  \bibinfo{author}{\bibfnamefont{J.}~\bibnamefont{{Lys}}},
  \bibinfo{author}{\bibfnamefont{H.}~\bibnamefont{{Murayama}}},
  \bibinfo{author}{\bibfnamefont{C.~G.} \bibnamefont{{Wohl}}},
  \bibnamefont{et~al.}, \bibinfo{journal}{\prd} \textbf{\bibinfo{volume}{86}},
  \bibinfo{eid}{010001} (\bibinfo{year}{2012}).

\bibitem[{\citenamefont{{Fraga} et~al.}(2001)\citenamefont{{Fraga}, {Pisarski},
  and {Schaffner-Bielich}}}]{Fraga2001}
\bibinfo{author}{\bibfnamefont{E.~S.} \bibnamefont{{Fraga}}},
  \bibinfo{author}{\bibfnamefont{R.~D.} \bibnamefont{{Pisarski}}},
  \bibnamefont{and}
  \bibinfo{author}{\bibfnamefont{J.}~\bibnamefont{{Schaffner-Bielich}}},
  \bibinfo{journal}{\prd} \textbf{\bibinfo{volume}{63}}, \bibinfo{eid}{121702}
  (\bibinfo{year}{2001}), \eprint{hep-ph/0101143}.

\bibitem[{\citenamefont{{Lora-Clavijo}
  et~al.}(2015)\citenamefont{{Lora-Clavijo}, {Cruz-Osorio}, and
  {Guzm{\'a}n}}}]{Lora2015}
\bibinfo{author}{\bibfnamefont{F.~D.} \bibnamefont{{Lora-Clavijo}}},
  \bibinfo{author}{\bibfnamefont{A.}~\bibnamefont{{Cruz-Osorio}}},
  \bibnamefont{and} \bibinfo{author}{\bibfnamefont{F.~S.}
  \bibnamefont{{Guzm{\'a}n}}}, \bibinfo{journal}{\apjs}
  \textbf{\bibinfo{volume}{218}}, \bibinfo{eid}{24} (\bibinfo{year}{2015}),
  \eprint{1408.5846}.

\bibitem[{\citenamefont{{Guzman} et~al.}(2012)\citenamefont{{Guzman},
  {Lora-Clavijo}, and {Morales}}}]{Guzman2012}
\bibinfo{author}{\bibfnamefont{F.~S.} \bibnamefont{{Guzman}}},
  \bibinfo{author}{\bibfnamefont{F.~D.} \bibnamefont{{Lora-Clavijo}}},
  \bibnamefont{and} \bibinfo{author}{\bibfnamefont{M.~D.}
  \bibnamefont{{Morales}}}, \bibinfo{journal}{ArXiv e-prints}
  (\bibinfo{year}{2012}), \eprint{1212.1421}.

\bibitem[{Note1()}]{Note1}
Note1, \bibinfo{note}{by taking the solutions of the Tolman-Oppenheimer-Volkoff
  equations given in section \ref {TOVEQ} reduce to the isotropic case when
  $(P-P_{\perp })=0$ in (Eq.(\ref {hyd}))}.

\bibitem[{\citenamefont{{The LIGO Scientific Collaboration} and {The Virgo
  Collaboration}}(2017)}]{Abbott2017}
\bibinfo{author}{\bibnamefont{{The LIGO Scientific Collaboration}}}
  \bibnamefont{and} \bibinfo{author}{\bibnamefont{{The Virgo Collaboration}}}
  (\bibinfo{collaboration}{LIGO Scientific Collaboration and Virgo
  Collaboration}), \bibinfo{journal}{Phys. Rev. Lett.}
  \textbf{\bibinfo{volume}{119}}, \bibinfo{pages}{161101}
  (\bibinfo{year}{2017}),
  \urlprefix\url{https://link.aps.org/doi/10.1103/PhysRevLett.119.161101}.

\bibitem[{\citenamefont{{Rezzolla} et~al.}(2018)\citenamefont{{Rezzolla},
  {Most}, and {Weih}}}]{Rezzolla2017}
\bibinfo{author}{\bibfnamefont{L.}~\bibnamefont{{Rezzolla}}},
  \bibinfo{author}{\bibfnamefont{E.~R.} \bibnamefont{{Most}}},
  \bibnamefont{and} \bibinfo{author}{\bibfnamefont{L.~R.}
  \bibnamefont{{Weih}}}, \bibinfo{journal}{Astrophys. J. Lett.}
  \textbf{\bibinfo{volume}{852}}, \bibinfo{eid}{L25} (\bibinfo{year}{2018}),
  \eprint{1711.00314}.

\bibitem[{\citenamefont{{Antoniadis} et~al.}(2013)\citenamefont{{Antoniadis},
  {Freire}, {Wex}, {Tauris}, {Lynch}, and {et al.}}}]{Antoniadis2013}
\bibinfo{author}{\bibfnamefont{J.}~\bibnamefont{{Antoniadis}}},
  \bibinfo{author}{\bibfnamefont{P.~C.~C.} \bibnamefont{{Freire}}},
  \bibinfo{author}{\bibfnamefont{N.}~\bibnamefont{{Wex}}},
  \bibinfo{author}{\bibfnamefont{T.~M.} \bibnamefont{{Tauris}}},
  \bibinfo{author}{\bibfnamefont{R.~S.} \bibnamefont{{Lynch}}},
  \bibnamefont{and} \bibinfo{author}{\bibnamefont{{et al.}}},
  \bibinfo{journal}{Science} \textbf{\bibinfo{volume}{340}},
  \bibinfo{pages}{448} (\bibinfo{year}{2013}), \eprint{1304.6875}.

\bibitem[{\citenamefont{Alford et~al.}(2005)\citenamefont{Alford, Braby, Paris,
  and Reddy}}]{alford2005hybrid}
\bibinfo{author}{\bibfnamefont{M.}~\bibnamefont{Alford}},
  \bibinfo{author}{\bibfnamefont{M.}~\bibnamefont{Braby}},
  \bibinfo{author}{\bibfnamefont{M.}~\bibnamefont{Paris}}, \bibnamefont{and}
  \bibinfo{author}{\bibfnamefont{S.}~\bibnamefont{Reddy}},
  \bibinfo{journal}{The Astrophysical Journal} \textbf{\bibinfo{volume}{629}},
  \bibinfo{pages}{969} (\bibinfo{year}{2005}).

\bibitem[{\citenamefont{Schmitt}(2010)}]{schmitt2010dense}
\bibinfo{author}{\bibfnamefont{A.}~\bibnamefont{Schmitt}},
  \emph{\bibinfo{title}{Dense matter in compact stars: A pedagogical
  introduction}}, vol. \bibinfo{volume}{811} (\bibinfo{publisher}{Springer},
  \bibinfo{year}{2010}).

\bibitem[{\citenamefont{{Rezzolla} and {Zanotti}}(2013)}]{Rezzolla_book:2013}
\bibinfo{author}{\bibfnamefont{L.}~\bibnamefont{{Rezzolla}}} \bibnamefont{and}
  \bibinfo{author}{\bibfnamefont{O.}~\bibnamefont{{Zanotti}}},
  \emph{\bibinfo{title}{Relativistic Hydrodynamics}}
  (\bibinfo{publisher}{Oxford University Press}, \bibinfo{address}{Oxford, UK},
  \bibinfo{year}{2013}), ISBN \bibinfo{isbn}{9780198528906},
  \urlprefix\url{https://books.google.com.co/books?id=KU2oAAAAQBAJ}.

\end{thebibliography}
%\bibliography{bibliography}               
%%%%%%%%%%%%%%%%%%%%%%%%%%%%%%%%%%%%%%%%%%%%%%%%%%%%%%%%%%%%%

\end{document}